\documentclass[preprint,epsf]{aastex}
\usepackage{epsfig}
\def\lsim{\lower0.6ex\vbox{\hbox{$ \buildrel{\textstyle 
<}\over{\sim}\ $}}}
\def\rsim{\lower0.6ex\vbox{\hbox{$ \buildrel{\textstyle 
>}\over{\sim}\ $}}}
\def\beq{\begin{equation}}
\def\eeq{\end{equation}}

\def\alwaysmath#1{{\ifmmode{#1}\else{$#1$}\fi}}
\def\he#1{\hbox{\alwaysmath{{}^{#1}}{\rm He}}}

\def\li#1{\hbox{\alwaysmath{{}^{#1}}{\rm Li}}}
\def\sun{${\,_\odot}$}

\def\hii{H\thinspace{$\scriptstyle{\rm II}$}~}
\def\etal{{\it et al.}~}
\begin{document}
\begin{flushright}
UMN-TH-1913/00\\
TPI-MINN-00/35\\
astro-ph/0007081 \\
June 2000
\end{flushright}
\vskip 0.75in
 
\begin{center} 

{\Large{\bf   On the Determination of the \he4 Abundance in 
Extragalactic \hii Regions}}
 
\vskip 0.5in
{  Keith~A.~Olive$^{1,2}$, and Evan~D.~Skillman$^2$  }
 
\vskip 0.25in
{\it

$^1${Theoretical Physics Institute, 
University of Minnesota, Minneapolis, MN 55455}\\
$^2${School of Physics and Astronomy,
University of Minnesota, Minneapolis, MN 55455}\\
}

%\newpage

\vskip .5in
{\bf Abstract}
\end{center}
\baselineskip=18pt 

Of the light element abundances capable of testing standard big bang
nucleosynthesis, only \he4 is measured with an accuracy of a few percent.
Thus, it is imperative to establish a comprehensive technique for
determining \he4 abundances and reliable estimates of the true
systematic uncertainties.  Helium abundance determinations in H II regions
are made from the observations of several distinct He I emission lines
and their strengths relative to H I emission lines.  With the general
availability of large format, linear CCD detectors, the accuracy of
relative emission line ratios has improved to the point where several
terms in the error budget which were assumed to be negligible may now be 
important.  Here we investigate the estimation of errors in deriving and
reporting nebular helium abundances from optical emission line spectra.

We first investigate the analysis of the H Balmer emission lines.
These lines can be used to determine the reddening of the spectrum,
but underlying stellar absorption needs to be accounted for.  Using 
a minimization routine, it is possible to solve simultaneously for
both reddening and underlying absorption, but, due to the degeneracy
of the sensitivities of the individual lines, a minimization routine
may underestimate the true errors in the solution.  Monte Carlo
modeling allows for a better estimate of the errors in underlying
absorption and reddening which need to be propagated to all of
the data.  The magnitude of the $\chi^2$ in such a minimization
is important in judging the reliability of the derived solution.
We also point out
that comparing corrected Balmer line ratios to their theoretical
values provides a sensitive test of the propriety of the magnitude
of the errors of reported emission line strengths.

The derived \he4 abundance depends on the H I and He I emissivities, 
the electron density, the electron temperature, the presence of 
underlying stellar absorption,
and, in some cases, the optical depth in the He I emission lines.
Certain He I emission lines depend sensitively on some of these quantities.
Ideally, solutions in which all observable He I lines yield the
same answer for the derived He abundance are favored. 
We examine several methods for such a ``self-consistent'' analysis to 
obtain the \he4 abundance in low metallicity \hii regions, and
attempt to make a thorough assessment of the uncertainties 
involved in such determinations.  
We demonstrate that solving for physical parameters via a
minimization routine opens up the possibilities of incorrect
solutions if there are any systematic problems with even one
observed He I emission line.  In many cases, minimizing with just
three lines ($\lambda$5876, $\lambda$4471, and $\lambda$6678)
is competitive with adding more lines into a minimization
(and always provides a useful diagnostic).  Underlying He~I
absorption can be important at the level of reported uncertainties,
yet hard to detect.  He~I $\lambda$4026 is shown to be a sensitive 
diagnostic of underlying He~I absorption, and we recommend adding 
it in to minimization methods.
We show that Monte-Carlo simulations
are necessary to reliably determine the uncertainties in
the physical parameters as well as the \he4 abundance 
determined in minimization routines.
We also point out that the magnitude of the $\chi^2$ is important
in judging the reliability of the derived solution and should be
reported in addition to the derived helium abundance.

\newpage
\baselineskip=18pt 
\noindent
\section{Introduction}

The only way to test big bang nucleosynthesis (BBN) and therefore 
cosmology at an age of order seconds to minutes, is through the 
observational abundances of the light elements D, \he3, \he4, and 
\li7 (see. e.g., Olive, Steigman, \& Walker 2000). Because there are no
measurements of
\he3 at very low  metallicity (i.e., significantly below solar) at this
time, a higher burden is placed on the remaining three elements.
The measurements of D/H in quasar absorption systems are 
very promising (Burles \& Tytler 1998a; 1998b), although not all data
agree (Webb et al., 1997; Levshakov, Tytler, \&
Burles 1998; Tytler \etal 1999). Similarly, \li7 measurements are
continually improving (Ryan, Norris \& Beers 1999) and systematic
uncertainties are being reduced (Ryan \etal 2000), but the accuracy of the
primordial abundance determinations for \li7 are not probably not much
less than a factor of 2.  Testing the theory of BBN  requires reliable
abundances of at least two isotopes.  
Unlike the other light element abundances, in order to be a useful
cosmological constraint, \he4 needs to be measured with a 
precision at the few percent level.   
Thus, the  determination of the \he4 abundance with improved accuracy 
continues to be of prime importance to cosmology. 

To date, the most useful \he4 abundance determinations are made by 
observing helium emission lines in HII regions of metal-poor dwarf 
galaxies. These measurements, which span metallicities ranging down 
to 1/50th of the solar oxygen abundance, all show \he4 abundances, 
$Y$, between 22 and 26\% by mass.  This is one of the strongest 
indications that the majority of the \he4 observed in these systems is 
in fact primordial and that BBN occurred.  At the next level of precision, 
however, it is necessary to be able to extract a primordial 
abundance, $Y_p,$ from these data (e.g., Pagel et al.\ 1992, 
hereafter PSTE). 
The most common method to determine 
$Y_p$ is by means of a linear regression with respect to a tracer 
element (Peimbert \& Torres-Peimbert 1974; 1976) such as oxygen or 
nitrogen (other methods such as a Bayesian analysis gives similar 
results, Hogan et al.\ 1997). To first order, we expect that along 
with the stellar production of heavy elements, there is a component 
of stellar contamination of primordial He. The uncertainty in the
primordial abundance of \he4 due to this contamination and its
exact relationship to the production of heavy elements is reduced 
by observing the lowest metallicity objects. 
Currently there is some controversy concerning the best estimate of
$Y_p$.  Izotov \& Thuan (1998b, hereafter IT98) assembled a sample of
45 low metallicity HII regions, observed and analyzed in a uniform
manner, and derived a value of $Y_p$ $=$ 0.244 $\pm$ 0.002 and
0.245 $\pm$ 0.001 (with regressions against O/H and N/H respectively).
This value is significantly higher than the value of $Y_p$ $=$ 0.228 
$\pm$ 0.005 derived by PSTE. 
Analysis based on the combined available data (Olive \& Steigman 1995;
Olive, Skillman, \& Steigman 1997; Fields \& Olive 1999) yield an
intermediate value of 0.238 $\pm$ 0.002 with an estimated systematic
uncertainty of 0.005.  Peimbert, Peimbert,
\& Ruiz (2000, hereafter PPR) have derived a very accurate helium
abundance for the HII region NGC 346 in the Small Magellanic Cloud, and
from this they infer a value of $Y_p$ $=$ 0.2345 $\pm$ 0.0026.  These
different results depend, in part, on differences in the analyses of the
observations.  Thus, it is important to better understand any
systematic effects that may result due to different analyses methods.
Furthermore, as one can plainly see, the differences in the various
determinations of the \he4 abundance appears to be many sigma. Thus it is
clear that present systematic errors have been underestimated, and the
main goal of this paper is to specify methods to better quantify and
reduce the systematic uncertainties in \he4 abundance determinations.
  
Of course, the degree to which we can make an accurate determination 
of the primordial He abundance ultimately depends on our ability 
to extract accurate \he4 abundances from individual extragalactic 
HII regions.  All of the information comes from the relative 
strengths of the emission lines of He I and H I, and the emission 
lines of heavier elements such as oxygen, nitrogen, and sulfur.
To determine a \he4 abundance from the emission line intensities, 
it is necessary to determine the physical characteristics of the 
HII region.  The electron temperature of the HII region is usually 
determined from the temperature sensitive ratio of [O III] 
emission lines (but see PPR).  
Electron densities can be derived from the ratios 
of [S II] and [O II] lines, although these may not be favored 
(see Izotov et al.\ 1999 and also section 2.4).  
While the relative H I and He I emissivities 
have very small dependencies on the electron densities, certain 
He I emission lines have an enhanced density dependence due to 
the collisional excitation of electrons out of the metastable 
2S state. Additionally, some He~I emission lines are subject to 
enhancement or diminuation through the radiative transfer effects 
of absorption or florescence.  

One also needs to ascertain whether or not neutral helium 
(or neutral hydrogen) corrections are important (e.g.,
Shields \& Searle 1978; 
Dinerstein \& Shields 1986; Viegas, Gruenwald, \& Steigman 2000). 
Finally, corrections due to possible effects of underlying He~I stellar 
absorption in the spectra must be considered, though in the past, 
these have usually been neglected. 
Underlying He I absorption was shown to be an important effect 
for the NW region of I Zw 18 (Izotov \& Thuan 1998a).  
Skillman, Terlevich, \& Terlevich (1998)
have demonstrated that the effects of underlying He I absorption
may be more important than claimed by IT98 and may explain some
of the ``anomalous'' line ratios observed by them (which led
to the rejection of certain objects from the linear
regressions used to determine $Y_p$).

In some studies (e.g., Skillman \& Kennicutt 1993; Skillman et al.\
1994), \he4 abundances determinations were based on a single emission 
line at $\lambda$6678. This line was deemed preferable as it is 
less subject to the effects of collisional enhancement relative 
to the stronger He I lines at $\lambda$4471 and $\lambda$5876
(cf. Pagel \& Simonson 1989). 
Its proximity to H$\alpha$ also means that the ratio 
$\lambda$6678/H$\alpha$ is practically unaffected by a reddening 
correction. 
However, there is always a danger relying on a single emission line. 
Fortunately, other He~I emission lines are available.  The three 
lines $\lambda$4471, $\lambda$5876,
and $\lambda$6678 are all relatively insensitive to density and 
optical depth effects. This means that, on the one hand, the 
conversion of these line strengths to a helium abundance can be 
done with greater certainty.  On the other hand, they do not 
provide a reliable estimate of the either the optical depth or 
density.  The latter is known to make a correction of order 
1\% to 5\% (depending on both the density and the temperature) 
due to collisional excitations. Nevertheless, one could make a 
case for using only these lines to determine the helium abundance. 

Recently, a ``self-consistent'' approach to determining the 
\he4 abundance was proposed by Izotov, Thuan, \& Lipovetsky 
(1994, 1997, hereafter ITL94, ITL97)
by considering the addition of other He lines.  First the 
addition of $\lambda$7065 was proposed as a density diagnostic, 
and then, $\lambda$3889 was later added to estimate the radiative 
transfer effects (since these are very important for $\lambda$7065).
This is the method used by IT98 in 
their most recent estimate of $Y_p$. 
While this method, in principle, represents an improvement 
over helium determinations using a single emission line, 
systematic effects become very important if the helium 
abundances derived from either $\lambda$3889 or $\lambda$7065 
deviate significantly from those derived from the other three 
lines (see \S 5). 

In this paper, we will attempt to better quantify the true 
uncertainties in the individual helium determinations in 
extragalactic HII regions. 
After a brief discussion of the available observables needed in 
the determination of $y^+ = $He$^+$/H$^+$, we present methods for 
determining the reddening correction, 
C(H$\beta$), the degree of underlying absorption in H~I and He~I, and
ultimately $y^+$. In addition we will show  specifically how various
uncertainties in measured quantities affect $y^+$. In section 4, we will
describe  several alternative methods for deriving the helium abundance 
based on 3 to 6 emission lines.
Here we will investigate 
the utility of adding a sixth He~I line, $\lambda$4026, which 
may be used to make a quantitative correction for the presence 
of underlying stellar  He~I absorption. 
For various cases based on several emission lines, we will test 
the minimization procedure (with respect to the recombination 
values) by calculating Monte Carlo realizations of the input data.  
This enables us to check the stability of any given solution of the
minimization and better estimate the uncertainties in our result. 
In section 5, we present some examples of synthetic data to 
demonstrate the power of the method and the dangers of 
systematic uncertainties in the observed He I line strengths. 
Our conclusions and prospects for accurate helium determinations 
will be given in section 6.

The goal of this paper is to explore different analysis methodologies
and to promote particular observational and analysis techniques.
In the future, we will apply the recommended methods to both new 
observations and other observations reported in the literature.

\section{Determination of Physical Parameters}

In this section, we will concentrate on the impact of the physical input
parameters on He abundance determinations.  We will discuss
the necessity of obtaining accurate line strengths and the limitations in
doing so. We pay special attention to reddening
as determined by H I line ratios. The uncertainties in this correction are
particularly important as they feed into the uncertainty in all of the
subsequent He I line strength determinations.  We will also discuss
determination of the electron temperature and density.

\subsection{Measurements of Relative Emission Line Strengths and Errors}

With the advent of large format, linear CCD array detectors in the
last decade, we are in the best position ever to obtain spectra of
emission line objects with the quality and accuracy necessary for
helium abundance measurements. 
While it may seem unnecessary to discuss the measurement of emission
line strengths here, this work starts with the assumption that
the spectra have been properly calibrated and that errors associated 
with that calibration have been taken into account.
Targets and standard stars should both be observed close to the
parallactic angle in order to minimize atmospheric differential 
refraction (Filippenko 1982). 
It is important to observe several standard stars (preferably
from the HST spectrophotometric standards of Oke 1990).  These 
standard stars are believed to be reliable to about 1\% across the
optical spectrum, and thus, this sets a fundamental minimum level
of uncertainty in any observed emission line ratio. Observations
of both red and blue stars allows a check on the possibility of second-order
contamination of the spectrum.
Typically, one-dimensional spectra are extracted from
long-slit (2-D) observations.  Special care needs to be taken setting
the extraction aperture width and the aperture should be sufficiently
wide that small alignment errors do not give rise to systematic errors
(this comes at a cost in signal/noise, but ensures photometric 
fidelity).  Given these potential uncertainties, it is unreasonable
to record errors of less than one percent in emission line ratios,
regardless of the total number of photons recorded.  Of course, multiple
independent measurements of the same target provide the best estimates
of true observational errors, and existing measurements of this type
confirm this minimum error estimate (Skillman et al.\ 1994).

It should also be noted that it is imperative to
integrate under the emission line profile (as
opposed to fitting the line with a Gaussian profile).  Fitting procedures
can introduce systematic differences between high signal/noise
and low signal/noise lines.  Given the dynamic range of the H~I and
He~I emission lines required to produce an accurate He/H abundance
(e.g., the faint He~I line $\lambda$6678 is about 1\% of H$\alpha$ and
He~I $\lambda$4026 is less than 2\% of H$\beta$), any systematic
error between measuring strong and faint lines will have dramatic
results. A special challenge is presented by the presence of underlying
stellar absorption.  The underlying absorption is generally broader
than the emission, so quite often, when observed at a resolution
of a few Angstroms or better, the H I or He I emission line is sitting in
an absorption trough.   Measuring all H I and He I emission lines in a 
consistent manner is important to obtaining a good solution for 
both the emission strength and the underlying absorption (see next
section).  Measurements at maximum resolution possible (while still
measuring all lines simultaneously) are preferred. 
  
\subsection{Determination of Reddening and Underlying H I Absorption 
from Balmer Lines}

Because (1) we know the theoretical emissivities of the recombination
lines of H I (e.g., Hummer \& Storey 1987),
(2) the ratios of the H I recombination lines in emission are relatively
insensitive to the physical conditions of the gas (i.e., electron 
temperature and density), and 
(3) there are a number of H I recombination lines
spread through the optical spectrum, it is possible to use the
observed line ratios to solve for the line-of-sight reddening of the 
spectrum (cf., Osterbrock 1989).
If one assumes a reddening law ($f$($\lambda$), e.g., Seaton 1979), in 
principle, it is possible to solve for
the extinction as a function of wavelength by measuring a single
pair of H I recombination lines. Values of C(H$\beta$),
the logarithmic reddening correction at H$\beta$, can be derived from:
\begin{equation}
log \left[ I(\lambda) \over I(H\beta) \right] =
  log \left[ F(\lambda) \over F(H\beta) \right] +
  C(H\beta)f(\lambda) , 
\label{chb}
\end{equation}
\noindent where $I(\lambda)$ is the intrinsic line intensity and
$F(\lambda)$ is the observed line flux corrected for atmospheric
extinction.  Assuming a reddening law introduces a degree of
uncertainty.  Studies in our Galaxy have shown that the reddening law
exhibits large variations between different lines of sight, but
these variations are most important in the ultraviolet
(Cardelli, Clayton, \& Mathis 1989).
Additionally, the total measured extinction can have both
Galactic and extragalactic components (and note the added complexity of
the shift in wavelength for the reddening law for systems at significant
redshift). Note that it is typical that no error is associated with the
assumption of a reddening law.  Davidson
\& Kinman (1985) point out that tying the He~I emission lines to the
nearest pair of bracketing H~I lines significantly reduces the impact of
assuming a reddening law (i.e., ``the interpolation advantage''), but
it is unlikely that there is absolutely no error incurred 
with this assumption.

Underlying stellar absorption will affect the ratios of individual
H I line pairs, so, in practice, it is best to measure
several H I recombination lines.   One can then solve for both
the reddening and the stellar absorption underlying the emission 
lines (e.g., Shields \& Searle 1978; Skillman 1985).  
It is generally assumed that the underlying absorption for the 
brightest Balmer H~I lines is constant in terms of equivalent width.  
It is not clear how much error is incurred through this assumption, 
and  inspection of stellar spectra shows that it is unlikely to
be true for the fainter Balmer emission lines (e.g., H8, H9, and higher).
However, one has the observational check of comparing these 
corrected lines to their theoretical values.

We recommend solving for the reddening and the underlying absorption
by minimizing the differences between the observed and theoretical
ratios for the three Balmer line ratios H$\alpha$/H$\beta$,
H$\gamma$/H$\beta$, and H$\delta$/H$\beta$.  Both H7 and H8 are
blended with other emission lines, so they cannot be used for this
purpose.  While the H9 and H10 lines are often not observed
with sufficient accuracy to constrain the reddening and absorption,
in high quality spectra, the relative strengths of
H9 and H10 provide a check on the derived solutions.
In Appendix A we describe our method of using a $\chi ^2$
minimization routine to determine the best values of C(H$\beta$),
the underlying equivalent width of hydrogen absorption ($a_{HI}$),
and their associated errors.

Figure 1 is presented for instructional purposes.
It shows a comparison of the observed and corrected
hydrogen Balmer emission line ratios for three synthetic cases.
In constructing this figure, synthetic
H I Balmer emission line spectra were calculated assuming
an electron temperature (18,000 K),
density (100 cm$^{-3}$), and H$\beta$ equivalent width (100 \AA ).
Balmer emission line ratios were derived
for three different combinations of reddening and absorption.
All emission lines and equivalent widths were given uncertainties
of 2\%. 
In the first case, the spectrum was calculated assuming reddening
and no underlying absorption.  The second case assumes underlying 
absorption and no reddening.  The third case has both. 
The open circles show the deviations of
the original synthesized spectra from the theoretical ratios
in terms of the synthesized uncertainties (2\% for all lines).
Note that reddening and underlying absorption induce corrections 
in the same direction for all three line ratios, i.e., the
H$\alpha$/H$\beta$ line ratio increases for increased reddening
and underlying absorption and the bluer Balmer line
ratios all decrease for both effects.  
This covariance results in a degeneracy, thereby decreasing the
diagnostic power of the corrections as we will show.

The filled circles in Figure 1 show the results of using the $\chi ^2$ 
minimization routine described in Appendix A. 
If such a minimization is used, then the $\chi ^2$ should 
be reported.  This allows one to make an independent check on
the validity of the magnitude of the emission line uncertainties.
As one can see, the minimization procedure accurately reproduces the
assumed input parameters.  In case 1, the minimization found C(H$\beta$)
$= 0.10 \pm 0.03$ and $a_{HI} = 0.00 \pm 0.57$. Similarly, for the other
two cases, we find C(H$\beta$) $= 0.00 \pm 0.03$, $a_{HI} = 2.00 \pm
0.59$ and C(H$\beta$)$= 0.10 \pm 0.03$, $a_{HI} = 2.00 \pm 0.59$
respectively. In all three cases, since the data are synthetic, the
$\chi^2$ values for the solutions are vanishingly small.  Appendix B
discusses cases from the literature where the $\chi^2$ values
are quite large, indicating either a problem with the original
spectrum, an underestimate of the emission line uncertainties,
or both.  

As a test to determine the appropriateness of the uncertainties 
for the values of C(H$\beta$) and $a_{HI}$ as produced by
the $\chi ^2$ minimization, we have run Monte Carlo
simulations of the hydrogen Balmer ratios.  The Monte Carlo
procedure is described in Appendix A.
Figure 2 shows the results of Monte Carlo simulations
of solutions for the reddening and underlying absorption from
hydrogen Balmer emission line ratios for three synthetic cases
based on the input parameters of case 3 of Figure 1.
That is, the original
input spectra had reddening with C(H$\beta$) $=$ 0.1 and
$a_{HI}$ = 2 \AA. For these values of 
C(H$\beta$) and $a_{HI}$, we have run the Monte Carlo for three 
choices of EW(H$\beta$) = 50, 100, and
200 \AA.  (EW(H$\beta$) = 100 was used in Figure 1). Let us first 
concentrate on the results shown in the bottom panel of Figure 2.  
Each small point is the minimization solution derived from a different 
realization of the same input spectrum
(with 2\% errors in both emission line flux and equivalent width).  
The large open point with error bars shows the mean result 
with 1$\sigma$ errors derived from the $\chi^2$ solution from 
the original synthetic spectrum.
The large filled point with error bars shows the mean result 
with 1$\sigma$ errors derived from the dispersion in the 
Monte Carlo solutions. 
Note that the covariance of the two parameters leads to error ellipses. 
The Monte Carlo simulations find the correct solutions, but the error
bars appropriate to these solutions are significantly larger than the
errors inferred from the single $\chi ^2$ minimization.
In this case there is a small offset in the mean solutions (mostly 
due to the fact that solutions with negative values are not allowed). 
In the bottom panel, the errors in C(H$\beta$) are 29\% larger and the
errors in $a_{HI}$ are about 61\% larger for the Monte Carlo simulations
compared to the single $\chi^2$ minimization.

The middle and top panels of Figure 2 show cases for decreasing 
emission line equivalent width.
Note that, given the input assumptions, the constraints on the 
underlying absorption are stronger in absolute terms for the lower 
emission line equivalent width cases.
In all three cases, the $\chi^2$ minimization errors are smaller
than those produced by the Monte Carlo simulations.  For the middle
panel, the differences are 41\% for C(H$\beta$) and 80\% for $a_{HI}$,
while for the top panel, the differences are 46\% for C(H$\beta$) and
86\% for $a_{HI}$.

These test cases have shown that the errors in C(H$\beta$) and the
underlying stellar absorption can be underestimated by simply using the
output from a $\chi^2$ minimization routine, and that Monte Carlo
simulations can be used to give a more realistic estimate of the
errors.
Based on this experience, we recommend that the best way 
to determine the true uncertainties in the derived
values of C(H$\beta$) and $a_{HI}$ is to run Monte Carlo
simulations of the hydrogen Balmer ratios.  Simply running
a $\chi ^2$ minimization will underestimate the uncertainty
(due, in large part, to the covariance of the two parameters being 
solved for).  Since the reddening correction must be applied to the
He I lines as well, this uncertainty will propagate into the final
estimation of the He abundance.  This uncertainty, we find, is 
too large to be ignored. 

 If He I lines are observed at a given wavelength $\lambda$,
their intensities relative to H$\beta$ after the reddening correction is
given by eq. (\ref{chb}). The ratios $I(\lambda)/I(H\beta)$ can then be
used self-consistently to determine the He abundance and the physical
parameters describing the HII regions, after the effects of collisional
excitation, florescence, and underlying absorption as described in the next
section. We can quantify the contribution to the overall He abundance
uncertainty due to the reddening correction by propagating the error in
eq. (\ref{chb}). Ignoring all other uncertainties in
$X_R(\lambda) = I(\lambda)/I(H\beta)$, we would write 
\beq
{\sigma_X \over X} = \ln 10~f(\lambda)~\sigma_{C(H\beta)}
\label{echb}
\eeq
In the examples discussed above, $\sigma_{C(H\beta)} \sim 0.04$ (from the
Monte Carlo), and values of $f$ are 0.237, 0.208, 0.109, -0.225, -0.345,
-0.396, for He lines at $\lambda \lambda$3889, 4026, 4471, 5876,
6678, 7065, respectively. For the bluer lines, this correction alone is 1
-- 2 \% and must be added in quadrature to any other observational errors
in $X_R$. For the redder lines, this uncertainty is 3 -- 4 \%.  This
represents the {\em minimum} uncertainty which must be included in the
individual He I emission line strengths relative to H$\beta$.
Note that these errors alone equal or exceed the 2\% errors in the
individual line strengths assumed for this exercise.
However, the magnitude of the reddening error
terms for the red lines can be reduced if these lines are compared 
directly to H$\alpha$. If the corrected H$\alpha$/H$\beta$ ratio is
identical to the theoretical ratio, then it would be allowable to
include only the uncertainty in the reddening difference between 
H$\alpha$ and the red He~I emission line.  On the other hand, it is 
frequently the case that the corrected H$\alpha$/H$\beta$ ratio is 
significantly different from the theoretical ratio.  

Finally, we should note that
an additional complication is the possibility that, in the highest
temperature (lowest metallicity) nebulae, the H$\alpha$
line may be collisionally enhanced (Davidson \& Kinman 1985;
Skillman \& Kennicutt 1993).  In their detailed modeling of
I Zw 18, Stasinska \& Schaerer (1999) have found this to 
be an important effect (of order 7\% enhancement in H$\alpha$).
If this is not accounted for, this has the effect of 
artificially increasing the determined reddening (and thus, 
artificially decreasing the helium abundances measured from
the lines redward of H$\beta$ (e.g., $\lambda\lambda$ 5876, 6678)
and increasing the helium abundances measured from lines blueward 
of H$\beta$ (e.g., $\lambda$4471).  More work along the lines
Stasinska \& Schaerer (1999) with photoionization modeling of
high temperature nebulae is needed to determine whether this
effect is common in these low metallicity regions.

\subsection{Electron Temperature Determinations from 
Collisionally Excited Lines}

Since the temperature is governed
by the balance between the heating and cooling processes, and since the
cooling is governed by different ionic species in different radial
zones, one expects different ions to have different mean
temperatures (cf.\ Stasi\'nska 1990; Garnett 1992).  While this is best
treated with a complete
photoionization model, a reasonable compromise is to treat the
spectrum as if it arose in two different temperature zones, roughly
corresponding to the O$^+$ and O$^{++}$ zones. Since the oxygen ions
play a dominant role in the cooling, this is a reasonable thing to do.
Deriving temperatures in the high ionization zone generally consists of
measuring the highly temperature sensitive ratio of the emission from the 
``auroral line" of [O III] ($\lambda$4363) relative 
to the emission from the ``nebular lines" of 
[O III] ($\lambda\lambda$4959,5007).
Temperatures for the low ionization zone are usually derived from
photoionization modeling (e.g., PSTE); although it is possible to derive 
temperatures in the low ionization zone from the [O~II] 
I($\lambda$7320 + $\lambda$7330)/I($\lambda$3726 + $\lambda$3729) ratio
and a similar ratio for [S~II] (e.g., Gonz\'alez-Delgado et al.\ 1994;
PPR).  

Note that, to date, usually only the temperature from the high ionization
zone is used to derive the He abundance, and the He which resides in
the low ionization zone is generally not dealt with in a 
self-consistent manner.  To estimate the potential size of this effect,
we can look at the data for SBS1159$+$545 from IT98.  In SBS1159$+$545,
19\% of the oxygen is in the O$^{+}$ state (and thus 81\% in the
O$^{++}$ state).  Assuming all of the gas to be at a temperature of
18,400 K (the [O III] temperature), a $\lambda$5876/H$\beta$ ratio
of 0.101 $\pm$ 0.002 yields a helium abundance of 0.0855 (before 
reduction to account for collisional enhancement and in agreement with
IT98).  Assuming 81\% of the gas to be at the [O III] temperature of
18,400 K and 19\% of the gas to be at the [O II] temperature of  
15,200 K results in a helium abundance of 0.0848, or a difference of
0.8\%.  While this is a small difference, it is not negligible when
compared to the reported uncertainty in the measurement. Curiously,
including the effects of collisional enhancement almost perfectly
cancels this effect for the reported density of 110 cm$^{-3}$
($y^+$ $=$ 0.0815 treated as a single temperature zone and 
$y^+$ $=$ 0.0811 treated as two temperature zones for this object).  
Thus, using a lower temperature for the $y^+$ in the O$^+$ zone can
increase or decrease the helium abundance depending on the density. The
main point here is that the temperatures used for the two zones  and the
helium abundance should be treated consistently  (as emphasized by PPR).

Steigman, Viegas, \& Gruenwald (1997) have investigated the effect of
internal temperature fluctuations on the derived helium abundances and
find this to be important in the high temperature regime.  
The presence of temperature fluctuations, when analyzed assuming no
temperature fluctuations, results in underestimating both the oxygen
and helium abundances (here only [S II] densities are used, which are
typically higher than the densities derived from He I lines).  Assuming  a
range of relatively large temperature fluctuations (with a maximum of
4000K) results in an overall shift in the derived primordial helium
abundance of about 3\%.  Steigman et al.\ have argued that,  in absence
of constraints on the temperature fluctuations, the  errors should be
increased to account for this uncertainty.

Peimbert, Peimbert, \& Ruiz (2000) have shown that the different
temperature dependences of the He~I emission lines can be used to
solve for the density, temperature, and helium abundance simultaneously
and self-consistently.  They point out that photoionization 
modeling consistently shows that the electron temperature derived
from the [O~III] lines is always an upper limit to the average temperature
for the He~I emission, and thus, assuming the [O~III] temperature
will always produce an upper limit to the true helium abundance.
Here we will not explore the possibility of adding the electron
temperature as a free parameter to our minimization routines.
This is, in part, because the main motivations are to explore the
method promoted by IT98, to explore the possibility of handling the
effects of underlying absorption, and also, because one of our 
main conclusions,
that Monte Carlo modeling is required for a true estimation of the
errors will be true regardless of the minimization parameters.
Nonetheless, this is a very important result with the implication
that most helium abundances reported in the literature to date are 
really {\it upper limits}.

\subsection{Electron Density Determinations from Collisionally Excited Lines}

    The average density can be derived  by measuring the relative
intensities of two collisionally excited lines which arise from a 
split upper level.  In the
``low density regime" collisional de-excitation is unimportant and
all excitations are followed by emission of a photon.  The ratio of
the fluxes then simply reflects the ratio of the statistical weights
of the two levels.  In the ``high density regime'', where the level
populations are held at the ratio of their statistical weights,
the emission ratio becomes the ratio of the product of the statistical
weights and the radiation transition probabilities.  In the intermediate
regime, near the ``critical density" the line ratios are excellent
density diagnostics.  The best known is that of [S II]
$\lambda$6717/$\lambda$6731 which is sensitive in the range from
10$^2$ to 10$^4$ cm$^{-3}$ and can be observed at moderate spectral
resolution.  At higher spectral resolution, one can use several other line
pairs (e.g., [O II] $\lambda$3726/$\lambda$3729).

In order to convert these line ratios into densities, one needs to
know the energy level separations, the statistical weights of the
levels, and the radiative and collisional excitation and de-excitation
rates.  Fortunately, one can use the five-level
atom program originally written by De Robertis, Dufour, \& Hunt (1987)
which has been made generally available within 
IRAF\footnote
{IRAF is distributed by the National Optical Astronomy Observatories,
which is operated by the Association of Universities for Research in Astronomy,
Inc.\ (AURA) under cooperative agreement with the National Science Foundation.}
by Shaw \& Dufour (1995).  
This program has the additional great advantage that the
authors have promised to keep the input atomic data updated.  

As emphasized by ITL94, ITL97 and IT98,
the [S II] line ratio suffers from two problems as a density
diagnostic: (1) it is measuring the density is the low ionization
zone, which may apply to less than 10\% of the emission in 
a low metallicity giant HII region, and (2) it is relatively 
insensitive to density below about 100 cm$^{-3}$.  Since the
collisional excitation of the He~I lines is important at the
1\% level down to densities as low as 10 cm$^{-3}$, the [S II] lines
are not ideal density indicators (cf.\ Izotov et al.\ 1999),
and deriving densities directly from the He~I lines is, in 
principle, preferable.  This is discussed further in \S 4.
However, calculating the density from the [S II] lines 
(and other collisionally excited lines) provides an excellent
consistency check on the density derived from the He I lines.
 
\section{Converting Individual He I Lines into He/H Abundances}

\subsection{He I and H I Emissivities}

   The F(He I)/F(H$\beta$) emission line flux ratios are converted
to intrinsic intensity ratios, I(He I)/I(H$\beta$), by correcting 
for reddening and underlying H~I absorption and then incorporating
the errors in these corrections (from eq. (\ref{echb})) into the errors in
the line ratios as discussed in section 2.2.  These intrinsic line ratios
can then be converted to  He/H abundance ratios by using the theoretical
emissivities  calculated from recombination and radiative cascade theory 
(e.g., Brocklehurst 1971; 1972). 
Here we use the H~I emissivities
calculated by Hummer \& Storey (1987) and the He~I emissivities
calculated by Smits (1996). See Appendix C for further details. Normally,
uncertainties in the H I and He I emissivities are not included in the error
calculations when determining He/H abundance ratios.  It is usually
assumed that these uncertainties are small in comparison with 
the other error terms, however, the quoted uncertainties on 
derived nebular helium abundances are becoming so small that this
assumption may no longer be true.
We would like to note that there is still a need for a modern
assessment of the uncertainties of the calculated He I emissivities.
Benjamin, Skillman, \& Smits (1999) have estimated that the 
uncertainty in the input atomic data alone may limit the 
accuracy to 1.5\%.

\subsection{Collisional Enhancement of He I Emission Lines}

At the high electron temperatures found in metal poor nebulae, 
collisional excitation from the metastable 2S level can become 
significant in determining the higher level 
populations in He I.   This effect has an exponential dependence on
electron temperature and a linear dependence on density.
Thus, the theoretical emissivities
need to be ``corrected'' for the radiative contribution of
these collisional excitations.  
In order to better calculate these collisional corrections to
the radiative cascade, quantum calculations of increasing accuracy
have been carried out to determine more exact collisional rates 
(Berrington et al.\ 1985; Berrington \& Kingston 1987; Sawey \& 
Berrington 1993).
Here we use the collisional rates of Sawey \& Berrington (1993)
and the resulting collisional corrections calculated by
Kingdon \& Ferland (1995).  In principle, it is better to 
join the collisional effects directly into the recombination
cascade calculation (e.g., as done by Benjamin et al.\ 1999),
but for the present exercise absolute numbers are less 
important than judging
the relative magnitudes of various effects.
One of the original motivations for this work was to reproduce the 
results published in IT98, so we have adopted an identical 
treatment of the input atomic data. 

\subsection{The Effects of Underlying Stellar Absorption}

A potential source of systematic error is the possibility of stellar 
absorption underlying the helium emission lines.  Certainly
there are typically many early type stars exciting the observed HII
regions, and certainly many of these stars have strong He I
absorption lines. 

Judging the 
degree to which underlying stellar absorption is important 
has been a real problem in the past (e.g., Shields \& Searle 1978).
Kunth \& Sargent (1983) proposed the very simple test of looking for
a trend in derived He abundance with EW(He I emission).
They found no evidence for this effect in their data (which span
approximately the same range in EW(He I emission) as modern day
observations).  Skillman, Terlevich, \& Terlevich (1998)
reexamined their data and found evidence for a slight trend in He/H 
with EW(H$\beta$) implying that underlying absorption may be 
present at a detectable level.  The theoretical modeling results of 
Olofsson (1995) have also been used as a guide in the past.  
These models indicated that the 
EW of $\lambda$4471 in absorption was generally of order
0.1 \AA\ or less.   However, Skillman, Terlevich, 
\& Terlevich (1998) pointed out that the model results may not be
representative of the typical extragalactic HII region observed
for these purposes, and that the
underlying absorption values may be much larger than 0.1.
They also drew attention to an inconsistency in the relative 
strengths of
He I absorption lines modeled by Olofsson.  That is, in observed 
stars (e.g., Lennon et al.\ 1993) and in numerical models (e.g., 
Auer \& Mihalas 1972), the strengths of the $\lambda$4471 and 
$\lambda$4026 lines are about a factor of two stronger than 
$\lambda$4387 and $\lambda$4922, while in the models of Olofsson, 
the opposite is true.  This potentially implies that the 
underlying absorption in $\lambda$4471 and $\lambda$4026
could have been underestimated by a factor of 4 in Olofsson's 
models (EWs for $\lambda$5876 and $\lambda$6678 are not calculated).
Revisiting the modeling by Olofsson with a view to the specific case
of determining nebular helium abundances remains a valuable
exercise for the future.

What are the greater implications for this realization that the effects
of underlying absorption could have been underestimated in the past?
Izotov \& Thuan (1998a) have demonstrated that underlying absorption is
important in the NW component of I~Zw~18.  IT98 
recognize the potential importance of underlying stellar
absorption.  They deal with this effect by (1) averaging over three
lines or (2) excluding a line from consideration when ``absorption
is evidently important''.

Here, we feel that a truly self-consistent approach will account 
for the effects of underlying absorption through detection and
correction for such effects.   
In the next section we present a method for doing this.
We pursue two different methods; first we include the 
possibility of underlying absorption in a $\chi ^2$ minimization
routine.  Second, we experiment with including a sixth line
$\lambda$4026, which has enhanced sensitivity to underlying
absorption.

In order to do this correctly, one must know, a priori, the
relative strengths of the underlying He I absorption lines.
We assume that the underlying He I stellar absorption lines
are all equal in terms of equivalent width.  Recall that we made a
similar assumption in the case of underlying H I absorption. Similarly, 
we cannot estimate how much systematic error we are incurring with this
assumption in the analysis of real observations.   However, by making the
same assumption in both the synthesized spectra and the analysis, we can
focus on the uncertainties in the method.  The assumption of identical
equivalent widths is probably not too bad.
Observations of individual Galactic B
supergiants (Lennon, Dufton, \& Fitzsimmons 1993) show that the
EW of the absorption lines of $\lambda$6678, $\lambda$4471,
and $\lambda$4026 are all of approximately equivalent strength
and share the same dependency on stellar effective temperature.
The models by Auer \& Mihalas (1972) show relatively good
agreement for EW($\lambda$4026), EW($\lambda$4471), 
EW($\lambda$5876), and
EW($\lambda$6678) for temperatures in excess of 35,000 K
and surface gravity values values of log g $=$ 4 and 4.5.

\subsection{He I Optical Depth Effects}

In order to compare observational measurements of helium
line intensities with theoretical values, it is necessary to 
consider radiative transfer effects and 
to determine what effects these have on the resulting line ratios.  
The standard references for radiative transfer in He~I emission 
lines are those of Robbins (1968)
and Robbins \& Bernat (1973). Recent examinations of this issue are
given by Almog \& Netzer (1989), Proga, Mikolajewska, 
\& Kenyon (1994) and
Sasselov \& Goldwirth (1995). Given the
improvements in the atomic data afforded by the re-examination of
A-values (Kono \& Hattori 1984), the recombination rates (Smits 1996), 
and collisional rates (Sawey \& Berrington 1993), 
a re-examination of radiative transfer
issues should be very useful. 
For the purpose of reproducing the IT98 results, here we will adopt the
fits given by IT98 to the modeling results of Robbins (1968) (the IT98\
equations are reproduced in Appendix C).   In Figure
3 we show the data from Robbins (1968) and the IT98 fits.  Note that
for the regime of low values of $\tau$(3889) relevant for the
current study (values of $\tau$(3889) $\ge$ 1.5 are rarely observed)
there is very little data available from Robbins (1968).
It is also important to note that these results represent only one
set of physical conditions.  An important parameter is the 
velocity gradient of the absorbing gas, which has been assumed to
be zero in the models chosen by IT98.
This is further motivation for a new study of the He~I radiative 
transfer effects.

\subsection{Ionization Correction Factors}

The degree to which the hydrogen and helium ionization zones in an
HII region coincide is generally determined by the hardness of the
ionizing radiation field, and may be governed, in part, by geometry
(e.g., Osterbrock 1989).  Thus, there is always concern that
in a specific observation of an HII region that neutral
helium is co-existent with ionized hydrogen along the line of sight
(see, e.g., discussion in Dinerstein \& Shields 1986).

Historically, a correction has been applied to the helium abundances
in order to correct for unobserved neutral helium.  V\'\i lchez \& Pagel
(1988), following the ideas of Mathis (1982), 
used the models of Stasi\'nska (1990) to demonstrate that
ratios of ionization fractions of sulfur and oxygen provided an
accurate measure of the hardness of the radiation field.  Pagel \etal
(1992) used this technique to determine whether such a
correction was necessary.  Their proposed methodology consisted of a
simple test: if the radiation field was soft enough that a significant
correction for neutral helium was implied, this correction was probably
too uncertain for the proposed candidate to be useful for a
helium abundance measurement.

ITL94 and ITL97 applied neutral helium corrections based on the models
of Stasi\'nska (1990), without adopting
the methodology of PSTE.  Unfortunately,
the correction derived in ITL94 is based only on the neutral helium
fraction and does not take into account the neutral hydrogen fraction
(see discussion in Skillman, Terlevich, \& Terlevich 1998).
IT98 revised these estimates assuming ionization
correction factors of one.

Viegas, Gruenwald, \& Steigman (2000) have produced photoionization
models indicating that H~II regions ionized by young, hot, metal-poor
stars may actually have more extended ionized helium regions when
compared to the ionized hydrogen.  This results in a ``reverse''
ionization correction, reducing the derived helium abundance by
as much as 1\% (cf.\ Figure 2 in Skillman, Terlevich, \& Terlevich
1998).  At present, lacking observational evidence of this effect, 
it is not clear that such a correction should be applied, but the
fact that it is of order the size of the errors presently quoted
on derived helium abundances implies that it should not be ignored
in the error budget.   

In this work we will simply assume that the ionization correction
factors are very close to one.  Skillman et al.\ (1994) noted the
constancy of He/H as a function of position in UGC 4483 despite 
significant variations in
oxygen ionization ratios. However, it is very difficult to 
constrain this uncertainty to less than 1\%.   

\section{Self-Consistent Methods for extracting the \he4 Abundance}

Having discussed many of the potential pitfalls in determining the 
\he4 abundance in individual extra-galactic HII regions, we can now
discuss the methodology for making such a determination.  As we noted
earlier, a He abundance can be inferred for each He~I emission line
observed by comparing the ratio of its observed intensity to H$\beta$
with the theoretical ratio and correcting for the effects of collisional
excitation, florescence and underlying He~I absorption. Thus, as per the
discussion of the previous sections, we need to determine three physical
parameters, the density, $n$, the optical depth, $\tau$, and the
equivalent width for underlying helium absorption, $a_{HeI}$. As argued by
ITL94 and ITL97, a self-consistent determination of the parameters, if
possible, is preferable.  Below we describe a few possible methods for
such a determination and stress the need for a careful accounting of the
resulting errors, which we deem requires a Monte Carlo simulation of the
data. 

As we noted above and discuss in detail below, different He~I lines are
more or less sensitive to the different physical parameters.
In principle, it is possible to fix these parameters by minimizing
$\chi^2$ using only the three best determined line strengths, $\lambda$4471, 
$\lambda$5876 and $\lambda$6678. 
However, because these line strengths are not very sensitive to
any of the physical parameters of interest, it may be preferable to
consider two or even more additional wavelengths. We describe these
various possibilities below. Once the parameters and their associated
uncertainties have been fixed, the He abundance may be determined by
averaging over all of the He~I lines used in the determination of the
physical parameters. 

We note that we are adopting a different philosophical approach
here compare to that in IT98.  In the final calculation of $y^+$, IT98
use only the main three lines to obtain the final He abundance.
Additionally, they adopt and report a minimum density of 10 cm$^{-3}$ 
(reduced from the minimum density of 50 cm$^{-3}$ adopted in ITL97)
and not lower densities which may be derived from their minimizations.
To be truly ``self-consistent'' would imply that the helium abundance is
derived from all observed lines and the physical parameters are those
resulting from the minimization.  An inspection of the IT98 data reveals
that often the He/H ratios derived from the $\lambda$7065 and
$\lambda$3889 lines are significantly different from the He/H ratios
derived from the main three lines. 
Additionally, when their minimization
routine is applied, one often finds unrealistically small values of the
density. We take these as warning signs that in some cases either the
minimization  is not finding the best possible solution due to a
degeneracy in the $\chi^2$ minimization, or there are problems with the
input data.
In such cases, it makes 
sense to either reject the object from derivations of the primordial
helium abundance or to attribute a larger uncertainty to account for
the lack of self-consistency in the minimization solution.
  
We begin our discussion of the merits of various minimization routines
by examining the dependence of the line strengths (for the
six He I lines of interest) on the physical parameters, $n, \tau,$ and
$a_{HeI}$.  Figure 4 shows six He~I emission lines and their relative 
dependences on the different effects discussed in the last sections.  We
show the relative effects for a baseline model of T $=$ 15,000K, 
$n =$ 10 cm$^{-3}$, $\tau$ $=$ 0, and no underlying stellar He I
absorption ($a_{HeI} = 0$).  The top panel of Figure 4 shows the effects
of an error of 500 K.  This is of order or larger than the errors
typically quoted for electron  temperatures for high quality spectra.  It
can be seen that reasonable errors in electron temperature (or
temperature fluctuations) will have a relatively small
effect on the derived He/H abundances (note, however, that Peimbert,
Peimbert, \& Ruiz 2000 have found that a coupling between temperature
and density allows solutions with small differences in temperature to
result in significant differences in density resulting in larger
than expected changes in the derived helium abundance).

The second panel from the top in Figure 4 shows the effect of
increasing the density from 10 to 100 and the subsequent 
collisional enhancement of the He I lines.  Clearly, of
the six lines, $\lambda$7065 is most sensitive to this effect.
Of the three lines normally used to calculate He/H abundances,
$\lambda$5876 is the most sensitive and $\lambda$6678 is
the least sensitive.  $\lambda$7065 would be an ideal 
density diagnostic if not for the sensitivity to 
optical depth shown in the third panel.

The third panel from the top in Figure 4 shows the effect of increasing 
the optical depth $\tau$(3889) from zero to one.  $\lambda$7065
has a strong sensitivity to optical depth effects.  $\lambda$3889
is also sensitive to $\tau$(3889), and in the opposite sense,
so that in combination these two lines could act to constrain
both density and optical depth.  Unfortunately, 
$\lambda$3889 is blended with H8 ($\lambda$3890).
Thus, in order to derive an accurate F($\lambda$3889)/F(H$\beta$)
ratio, the F($\lambda$3890) must be subtracted off and
underlying stellar H~I (and He~I) absorption must be corrected for.
This generally implies a relatively large uncertainty for 
F($\lambda$3889), and thus, a larger uncertainty in
the density and optical depth measures than one would
hope for.

Finally, the bottom panel in Figure 4 shows the effects of 
0.2 \AA\, of underlying stellar absorption.  The difference
of a factor of three between the effect on $\lambda$5876
and $\lambda$4471 and $\lambda$6678 means that there is
some sensitivity to underlying absorption through the
analysis of just those three lines.  However, 
the effect is very strong for the weaker $\lambda$4026 line.
Thus, we will explore the possibility of adding $\lambda$4026
as a diagnostic line.

It is very important to note from Figure 4 the strong trade-off
between density and underlying He~I absorption.  All six He~I 
line strengths are increased by increasing the density, while all
six He~I line strengths are decreased by increasing the underlying
He~I absorption.  While the relative effects vary from line to line,
the main result is a basic trade-off between density and underlying
absorption when both are included in a minimization routine.  This
means that adding underlying absorption as a free parameter in a
minimization routine will open up a larger range of parameter space
for good solutions.  On the other hand, it means that if absorption
is not included in minimizations, its effects may be masked by
driving the solutions to lower He abundances or densities.

\subsection{Using  3 Lines}

In principle, under the assumption of small values for the optical
depth $\tau$(3889), it is possible to use only the three bright lines
$\lambda$4471, $\lambda$5876, and $\lambda$6678 and still solve 
self-consistently for He/H, density, and $a_{HeI}$.
Of course, because these lines have relatively low sensitivities 
to collisional enhancement, the derived uncertainties 
in density will be large. However, as we will show, if there is some
reason to suspect a problem with any of the additional lines, the three
line method can actually lead to a more accurate result, and hence
should be used as a diagnostic if nothing else.  Using a 
minimization routine, as opposed to a direct solution, it is not 
even necessary to assume that $\tau$(3889) $=$ 0 in order to
derive a helium abundance from just three lines. 

The detailed procedure we use to determine the physical parameters
along with the He abundance is given in Appendix C. The procedure is
actually independent of the number of lines used, though when using fewer
lines (as in the present case of 3 lines) the results are likely to be
less robust.  

\subsection{Using  5 Lines}

A self-consistent approach to determining the
\he4 abundance was proposed by Izotov, Thuan, \& Lipovetsky
(1994,1997)
by considering the addition of other \he\,  lines.  First the
addition of $\lambda$7065 was proposed as a density diagnostic
and then, $\lambda$3889 was later added to estimate the radiative
transfer effects (since these are important for $\lambda$7065).
By minimizing the difference between
the ratios of $\lambda$3889/$\lambda$4471, $\lambda$5876/$\lambda$4471,
$\lambda$6678/$\lambda$4471, and $\lambda$7065/$\lambda$4471 and their
recombination values, the density, optical depth, and helium
abundance can be determined. The latter is determined by a
weighted mean of the helium abundance based on $\lambda$4471,
$\lambda$5876, $\lambda$6678 once the values of $n$ and $\tau$(3889)
are fixed.  This is the method used by IT98
in their most recent estimate of $Y_p$.
Underlying He~I absorption is assumed to be negligible in
their method.

While this method, in principle, represents an improvement
over helium determinations using a single emission line,
systematic effects become very important if the helium
abundances derived from either $\lambda$3889 or $\lambda$7065
deviate significantly from those derived from the other three
lines. In addition, working with the ratios of all of the
He I lines to a single He I line puts
undue weight on that single line (in this case $\lambda$4471).
This is especially vulnerable to systematic errors in the
presence of undetected underlying stellar absorption.

Here, we also consider using these five lines for determining the He 
abundance along with the physical parameters.  However, as described in the
appendix B, our minimization procedure is based on the weighted average
of the He abundance as determined from the five lines {\it independently}. 
We allow for the presence of underlying He~I absorption through the
assumption that it will be identical (in terms of equivalent width) for
all of the He~I lines.
In addition, once the physical parameters have been determined by 
the minimization, all five values of $y^+(\lambda)$ are used in a 
weighted mean to determine the final $Y^+$.

\subsection{Using  6 Lines}

Adding $\lambda$4026 as a diagnostic line increases the leverage
on detecting underlying stellar absorption.  This is because
the $\lambda$4026 line is a relatively weak line. However, this also
requires that the input spectrum is a very high quality one.
$\lambda$4026 is also provides exceptional leverage to underlying
stellar absorption because it is a singlet line and therefore
has very low sensitivity to collisional enhancement (i.e., $n$)
and optical depth (i.e., $\tau$(3889)) effects.

Our procedure for this case is identical to the one above with the
addition of the sixth line.  By adding $\lambda$4026 as a diagnostic
line, we increase our dependence on the assumption of equal equivalent
width of underlying absorption for all of the He~I lines.  
Our philosophy is that it is most important to discover underlying
absorption when it is present. 
If underlying absorption is important in an individual spectrum,
conservatively, it may be better to reject the object from 
consideration from studies constraining the primordial helium 
abundance.
If a solution implies significant underlying absorption, and all
of the helium lines give the same abundance within errors, it may
be taken as an endorsement of the assumption of equal EW of
underlying He~I absorption.

\section{Test Cases and Examples}

In this section we present the results from a number of test cases
varying the input physical parameters, the number of He~I emission 
lines used in the minimizations, and assumptions about certain
physical parameters.  Our philosophy here has been to test for 
relatively small variations, since the final goal is helium 
abundances for individual nebulae with accuracies approaching 
1\%.  That is, we are confident that if assumptions are grossly in
error that the derived abundances are wrong, but, more importantly,
if there is a very subtle effect (e.g., a very small amount of
underlying absorption or a small amount of optical depth), we need 
to understand how that will affect our derived helium abundances.

\subsection{Cases with no Systematic Errors}

We present here the results of running a few series of test 
cases.  In all cases, input spectra were synthesized with 
the prescriptions and assumptions described above or in the
appendices.  We chose a baseline model of T $= 18,000 \pm 200$ K,
EW(H$\beta$ $=$ 100), and He/H $=$ 0.080.  We then varied
the density, $a_{HeI}$, and $\tau$(3889) to produce different
cases.  Errors of
2\% were assumed for all of the input emission lines and
equivalent widths, and then each of these models were run 
through Monte Carlo realizations.  We then analyze the
resulting distributions of the results from a $\chi ^2$
minimization solution for He/H, density, $a_{HeI}$,
and $\tau$(3889).  

Figure 5 presents the 
results of modeling of 6 synthetic He I
line observations for a single case.  The  
four panels show the results of a density = 10 cm$^{-3}$,
$a_{HeI}$ = 0, and $\tau$(3889) $=$ 0 model. 
The solid lines show the input values (e.g., He/H = 0.080)
for the original calculated spectrum.  The solid circles
(with error bars) show
the results of the $\chi ^2$ minimization solution 
(with calculated  errors) for
the original synthetic input spectrum.  
The small points show the results of Monte Carlo realizations
of the original input spectrum.
The solid squares (with error bars) show the means and dispersions
of the output values for the $\chi ^2$ minimization solutions of
the Monte Carlo realizations.

Figure 5 demonstrates several important points.  First,
our $\chi ^2$ minimization solution finds the correct input
parameters with errors in He/H of about 1\% (less than the
2\% errors assumed on the input data, showing the power of
using multiple lines).  In this low density
case, the Monte Carlo results are in relatively good agreement with the
input data, with similar sized error bars.  There is a small
offset to lower densities and a similar small offset to non-zero
values of underlying absorption.  We found this effect throughout
our modeling, that when an input parameter such as underlying
He~I absorption or $\tau$(3889) is set to zero, the minimization
models of the Monte Carlo realizations (cases with errors) always
found slightly non-zero values (although consistent with zero)
in minimizing the $\chi^2$.  Note that in the lower right panel of
Figure 5 that the values of the $\chi ^2$ do not correlate 
with the values of $y^+$.  The solutions at higher values of 
absorption and $y^+$ are equally valid as those at lower
absorption and $y^+$.

Figure 6 presents the results of modeling of 6 synthetic He I
line observations for a case identical to that of Figure 5
with the exception of a higher density of 100 cm$^{-3}$.
For the $n = 100$ cm$^{-3}$ case, there is a systematic trend for the
Monte Carlo realizations to tend toward higher values of
He/H.  This is because, again, the inclusion of errors has allowed
minimizations which find lower values of the density and
non-zero values of underlying absorption and optical depth.
However, in this case, there is more ``distance'' from 
the lower bound of $n=0$, and thus more parameter space to allow the
effects of the parameter degeneracy to be noticed.  Note that the size of
the error bars in He/H have expanded by roughly 50\% as a result.  We can
conclude from this that simply adding additional lines or physical
parameters in the  minimization does not necessarily lead to the correct
results. In order to use the minimization routines effectively, one must
understand the role of the interdependencies of the individual
lines on the different physical parameters.  Here we have shown
that trade-offs in underlying absorption and optical depth allow
for good solutions at densities which are too low and resulting
in helium abundance determinations which are too high.
{\it This is one of the central results of this study.}
Again, note that there is no trend in the values of $\chi ^2$
with $y^+$.

Table 1 summarizes the results of a number of different test cases
like those shown in Figures 5 and 6.
Table 1 is grouped into six different cases of input with five 
different minimization routines.
The first two cases correspond those shown in Figures 5 and 6.
The other four cases consider non-zero values of underlying 
absorption, $\tau$(3889), or both.  The first two columns show
the results of minimizing on 3 lines (both assuming $\tau$(3889)
is zero and solving for $\tau$(3889)). The next two columns show
the results of minimizing on 5 lines (both assuming 
zero underlying absorption and solving for the underlying 
absorption).  The last column shows the results of the six line 
method which was used to produce Figures 5 and 6.  

The numbers in the table correspond to the average of the Monte 
Carlo results and their dispersion. The row labeled He/H gives
the results from averaging the He/H abundances from all of the
lines (3, 5 or 6), while the ``He/H (3)'' row gives the He/H values
derived from averaging only the three main He~I lines (after solving
for the physical parameters).
Note that the straight minimizations of the input data always returned
the input data (except in the cases where an assumption is inconsistent
with the input data).  Deviations of the Monte Carlo solutions from
the input values result because of: (1) inconsistencies between the
input data and input assumptions, (2) asymmetries in the Monte Carlo
distributions (e.g., in Figure 5, because the absorption is not allowed
to go negative, the distribution is truncated on one side, and thus
there is a bias to higher values of $y^+$), (3) degeneracies between
different parameters which result in lower $\chi ^2$ values for
values of the physical parameters quite far from their input values.

For the first two cases, (no underlying absorption and no optical 
depth), as expected, the 3 line method constraints
on the density are not strong.  However the derived helium abundances
are consistent, within the errors, with the input values.  
For the 5 line method, since the first two cases (1 and 2) have no 
underlying absorption, the 
method with the correct assumption finds a solution much closer to
the correct result (although all solutions are consistent with the
correct result, within errors).  Again, it is the degeneracy between 
density and underlying absorption which is responsible for 
the derived low density and high He abundance.  
Note, interestingly, that the three line method did not do
any worse (in fact it did slightly better) than the 5 line methods,
unless we assume a priori the correct answer for underlying 
absorption ($a_{HeI}=0$). Similarly, assuming $\tau =
0$ also improves the result in this case for obvious reasons.
The six line method, within the errors, gave results
consistent with but not equal to the input parameters.  Indeed, there is a
systematic trend to lower density and some underlying absorption even
when there is none.  The net result is a higher estimate of the He
abundance. This systematic trend can be traced to the degeneracy in the 
trends imposed by the different input parameters.
However, the 6 line method does significantly better than the 5 line 
method at constraining the underlying absorption (as
the $\lambda$4206 line anchors the values of $a_{HeI}$).

We learn more about the various methods when we consider the remaining
cases in Table 1. When $\tau(3889) \ne 0 $ (and $a_{HeI} = 0$), the five
and six line methods give very accurate results although, once again,
$\lambda 4206$ is needed to pin down the value of $a_{HeI}$ and break the
degeneracy. When the input value of $a_{HeI} \ne 0$, then only the 5 line
method which assumes $a_{HeI} = 0$ does badly.  The 5 and 6 line methods
which solve for $a_{HeI}$ do quite well. Figure 7
shows the results of the Monte Carlo when both $\tau$ and $a_{HeI} \ne
0$, and $n = 100$ cm$^{-3}$, i.e., case 6 of Table 1.
Thus it is encouraging that in perhaps more realistic cases where the input
parameters are non-zero, we are able to derive results very close to their
correct values. 

Indeed, Figure 7 shows many of effects we have been describing in the
previous cases. The average of Monte Carlo realizations is remarkably
close to the straight minimization for all of the derived parameters ($n,
a_{HeI}, \tau$ and $y^+$). However, there is an enormous dispersion in
these results due to the degeneracy in the solutions with respect to the
physical input parameters.  This results in error estimates for
parameters which are significantly larger than in the straight
minimization.  For example, the uncertainties in both the density and
optical depth are almost a factor of 3 times larger in the Monte Carlo. 
When propagated into the uncertainty in the derived value for the He
abundance, we find that the uncertainty in the Monte Carlo result (which
we argue is a better, not merely more conservative, value) is a factor of
2.5 times the uncertainty obtained from a straight minimization using 6
line He lines.  This amounts to an approximately 4\% uncertainty in the He
abundance, despite the fact that we assumed (in the synthetic data) 2\%
uncertainties in the input line strengths.   This is an unavoidable
consequence of the method - the Monte Carlo routine explores the 
degeneracies of the solutions and reveals the larger errors that
should be associated with the solutions.

{}From the above, we conclude (1) that adding absorption to the
minimization routines can lead to much larger regions of valid solution
space; (2) the trivial result that assuming no underlying absorption will
lead  to incorrect solutions in the presence of underlying absorption, 
(3) that adding an accurate $\lambda$4026 observation to a minimization 
solution will provide strong diagnostic power for underlying stellar 
absorption, and (4) that Monte Carlo models are required to determine the 
true uncertainties in the minimization results.

\subsection{Cases with Systematic Errors in I($\lambda$3889)}

In \S 4, it was shown that $\lambda$3889 is strongly sensitive to
optical depths effects and is required if $\lambda$7065 is to
be a good tracer of density.  It was also pointed out that,
unfortunately, $\lambda$3889 is blended with H8 ($\lambda$3890).
Thus, in order to derive an accurate F($\lambda$3889)/F(H$\beta$)
ratio, the F($\lambda$3890) must be subtracted off and
underlying stellar H~I (and He~I) absorption must be corrected for.

In the methodology of IT98, the contribution to He I $\lambda$3889
from H I emission is subtracted off by assuming the theoretical
value for the H I emission (typically, the He I emission accounts 
for almost 50\% of the blended line).  The total emission has to
be corrected for underlying stellar absorption, which is assumed to
be a constant equivalent width for all of the H I lines.  This
assumption is a potentially dangerous one.  Spectral studies of 
individual stars show that while this may be a good assumption to
first order, the equivalent widths of the higher order Balmer lines
are not strictly identical (see, for example, the spectral atlas
of Galactic B supergiants of Lennon, Dufton, \& Fitzsimmons 1992).
Secondly, this correction is usually large (corrected He I 
$\lambda$3889 emission line equivalent widths generally lie in
the range of 4 to 10\AA\  compared to the underlying absorption 
which is in the range of 0.5 to 3\AA ).  A good test of the
uncertainty in this correction would be a comparison of the
corrected higher order (H9 and H10) H I emission line strengths
compared to their theoretical values.  In Appendix B, we show
a few cases in the literature, where these comparisons reveal
evidence of a problem.

Here we investigate the possible effects of a systematically
low strength of $\lambda$3889 motivated by the possibility of
oversubtracting the underlying H I absorption.  We have run 
identical cases and analyses as in Table 1, but altered the input 
synthetic spectra by decreasing the relative flux and equivalent width
of $\lambda$3889 by 10\%. These results are presented in Table 2.
The first two columns are identical (since they are based on
only three unaffected lines) and are repeated for comparison.
Note that this exercise was motivated, in part, by the
systematically low values of He/H derived from $\lambda$3889
when compared with the main three He~I lines in a subsample of
the highest quality data from IT98.

The 10\% drop in $\lambda$3889 has dramatic effects. 
From \S 4, and especially Figure 4, it can be seen, that an 
underestimate of I($\lambda$3889) will lead to both artificially high
values of $\tau$(3889) and artificially low values of the density.
This is born out in inspection of Table 2. Beginning with the
cases in which the input values of $\tau$ and $a_{HeI}$ are 0, we see
that the density is grossly underestimated in both input cases with $n=
10$ and 100 cm$^{-3}$. In the low density case, the He/H solutions
based on all available lines are low, while the He/H(3) solutions
are close to correct.  This main effect is due to including the
low value of He/H from $\lambda$3889.  However, in the high density
case, it is the values of He/H(3) which are in error on the high
side.  This is due to the underestimate of the density (and thus
the corrections for collisional enhancement are too small).
Note that the solutions for $\tau$(3889) have been driven to
large values.  For both the higher density cases, the density
has been underestimated, the $\tau$(3889) overestimated, and
the He/H(3) overestimated. 

In Figure 8, we show the results of the 6 line method Monte 
Carlo for case 2 of Table 2.  Here the low density, high 
$\tau$(3889), and bias towards higher values of He/H are
clear (even higher values would be shown if He/H(3) were plotted).
Note that the lower right panel shows that the $\chi ^2$ values
are all systematically high for this case.  It would appear
that the $\chi ^2$ is a sensitive test to check whether
the I($\lambda$3889) values are systematically biased.

The low density cases show 
generally low values of He/H and satisfactory values of
He/H(3) (although the solutions for $\tau$(3889) are all
systematically high).  The main result of a systematically 
low values of I($\lambda$3889) occurs when the nebula has a 
high enough density that the collisional enhancement is 
important. Then the underestimate of density results in
systematically higher He/H. 

\subsection{Cases with Systematic Errors in I($\lambda$7065)}

One of the possible problems of using $\lambda$7065 is that in
a spectrum where the entire wavelength range from $\lambda$3727 to
$\lambda$7065 is observed in first order, there is the potential
for contamination of the red part of the spectrum from blue light
in the second order.  In principle, if a blue cutoff
filter is used (e.g., a CG385 order separation filter) then there
should be little contamination blueward of 2 $\times$ 3850 \AA\,
or 7700 \AA .  However, order separation filters are not 
perfect cut-on filters at 3850 \AA, but rather start to filter
out light above 4000 \AA , reach 50\% transparency at about 
3850 \AA\  and then drop to zero transparency somewhere between
3500 and 3600 \AA.  Thus, there is potential for some second order
contamination for all wavelengths above 7000 \AA .  If the 
observed target has a significant redshift, then the He~I 
$\lambda$7065 is even more susceptible to this problem.  The problem
is worse for bluer order separation filters (e.g., CG375). 

This effect can be quite subtle and there are two separate problems 
to consider.  The first is contamination of the standard star spectrum. 
If a blue standard star (e.g., a white dwarf) is used for flux
calibration, then the far red part of the spectrum detects 
additional second order blue photons, which, when used to 
calibrate the spectrograph, results in an overestimate of
the red sensitivity of the spectrograph.  The second problem
occurs in the target spectrum.  Here the far red continuum will
be contaminated by extra second order blue photons.  It is 
possible that if the blue spectral shape of the standard star
is similar to the spectral shape of the target, then these 
two effects will compensate, giving a rather normal looking
red continuum.  However, the overestimate of the red sensitivity
will result in underestimated emission line fluxes and equivalent
widths.  Since $\lambda$7065 lies right at the border of where
this effect can become important, it is very important to check
for this possibility.  This is most easily done by obtaining
spectra of both red and blue standard stars and deriving 
instrument sensitivity curves independently for the two stars.
The wavelength at which the two sensitivity curves begin to 
deviate indicates the onset of second order contamination.

Here we investigate the possible effects of a systematically
low strength of $\lambda$7065.  We have run identical cases 
and analyses as in Table 1, but altered the input synthetic 
spectra by decreasing the relative flux and equivalent width 
of $\lambda$7065 by 5\%. These results are presented in Table 3.
Note that this exercise was motivated, in part, by the 
systematically low values of He/H derived from $\lambda$7065 
when compared with the main three He~I lines in a subsample of
the highest quality data of IT98.  

The 5\% drop in $\lambda$7065 has dramatic effects. Beginning with the
cases in which the input values of $\tau$ and $a_{HeI}$ are 0, we see
that the density is grossly underestimated in both input cases with $n=
10$ and 100 cm$^{-3}$. In the low density cases, 
this does not have a very strong effect on the derived helium abundances
(because the density dependent collisional enhancement term is already
quite small).

In the high density case with no underlying absorption or optical
depth effects, the five and six line methods give a He abundances
which are significantly higher than the correct value of 0.08. In this
case, the three line method which is not distracted by the errant
$\lambda$7065 line does the best at finding the solution. We
see the effect of the $\lambda$7065 line driving down the density and
compensating by allowing for non-zero absorption, which is controlled in
the 6 line method due to $\lambda$4026.  

In Figure 9, we show the results of the six-line method Monte Carlo for 
case 2 of Table 3.  Case 2 shows the most discrepant results in Table 3,
and the six-line method is only better than the five-line method with
absorption allowed (but not constrained by $\lambda$4026.  Here it 
can be seen clearly that the solutions all favor lower density, and
thus higher He/H.  The trade-off between low density and high values
of underlying absorption is clearly shown.  Interestingly, the values
for $\chi^2$ are relatively low and quite satisfactory.  Clearly the
$\chi ^2$ is not a good diagnostic of an underestimated 
I($\lambda$7065), as the degeneracies allow mathematically 
acceptable solutions.

Case 6 provides an interesting comparison case to consider when 
$\lambda$7065 is low.
Notice here, that the 5 and 6 line methods again over-estimate the He
abundance. While both solutions find the approximate underlying
absorption (the 6 line method is better) they underestimate the density
and the optical depth.  Here the 3 line method, even with its lack of
sensitivity, still solves for the correct density (to within 8\% when
$\tau$ is not assumed to be zero) and underlying absorption.  The He
abundance is again very accurately determined by this method. 

In Figure 10, we show the results of the six-line method Monte Carlo 
for case 6 of Table 3.  In this case of a small amount of optical
depth and a small amount of underlying absorption, the solutions do a 
pretty good job of finding the correct range of density, optical depth, 
and underlying absorption.  The underestimated $\lambda$7065 has resulted
in a bias toward lower density, which has resulted in a bias toward
larger He/H, but the effect is not very large.  The main effect is
the size of the error bars.  Here it is clear that Monte Carlo errors 
in He/H are about 3 times larger than the errors from a straight
minimization.   Note again that the $\chi^2$ values are generally
small.

The main result of tests with a systematically low $\lambda$7065 
is that for nebulae with densities which correspond to significant
collisional enhancement corrections, the He/H will be overestimated.
The $\chi^2$ is not necessarily a good test of whether $\lambda$7065
has been systematically underestimated.

\section{Discussion}

We have pointed out a number of points in the analysis of optical
spectra of nebulae that can lead to biased or inaccurate abundance
results if not accounted for properly.  We hope to draw attention
to the treatment of the H~I Balmer lines as a critically important
step in an accurate abundance analysis. 

Our inspection of self consistent minimization methods for 
determining helium abundances has revealed several things.
In the perfect world when all of the uncertainties in the input data are
under control, the 6 line method is best at solving for the physical
input parameters and ultimately the He abundance. 
However, minimization routines should used with caution with
a eye toward systematically biased observations.
The key
diagnostics are the values of the $\chi^2$ in a straight minimization,
and the results of the 3 line method.  When the data are good the
$\chi^2$ per degree of freedom should be small, and the results of the 3,
5, and 6 line methods should be consistent. The latter methods should be
more accurate and carry smaller error bars on the derived quantities.
If either of these conditions are not met, then it is probably not
advisable to use those data in a determination of the primordial He
abundance.  In any case, a Monte Carlo realization of the observations
should be conducted to assess the true uncertainties in the resultant
abundances.

While it may seem that, in some cases, the straight minimization gives a
solution closer to the original input parameters than does the average
Monte Carlo result, one must bear in mind  that we have been using
synthetic data with known values.  By running the Monte Carlo on
synthetic data we are modeling possible sets of observations consistent
with the true physical description of the HII region.  That is, each Monte
Carlo minimization represents the result of a single possible
observation.  Running a Monte Carlo realization of real data
allows an exploration of all of the possible suitable solutions
allowed by the degeneracies in the sensitivities of the various
He~I lines to the different physical parameters.

Clearly the abundances used in estimating $Y_p$
are not observed but rather derived quantities. As we have seen, the
derivation of the He abundance relies on several, a priori, unknown but
physical input parameters. In this paper, we hope to have clarified the
determination process, and quantified the uncertainties in the result. 
Thus, it may be premature to be arguing over the
3rd decimal place in $Y_p$ until a systematic treatment and Monte Carlo
analysis of the data has been performed. 
Our purpose here is not to propose a minimum error for all nebular helium 
abundances nor to try to give a quantitative estimate of how a given
analysis may result in a systematic bias in the derivation of $Y_p$.
Rather, we wish to promote a methodology
for the analysis of all nebular HII region spectra in the pursuit of
accurate He abundances. We emphasize the importance of reporting more 
information (the equivalent widths of all of the H~I and He~I emission 
lines, the $\chi^2$ results for minimizations) and the use of Monte
Carlo techniques for characterizing error terms.  In the future, we
will apply the recommended methods to both new observations and 
other observations reported in the literature.

\vskip 0.5truecm
\noindent {\bf Acknowledgments}
\vskip 0.5truecm
We would like to thank R. Kennicutt
and B. Pagel for insightful comments on the manuscript. We also are
pleased to thank R. Benjamin, D. Kunth, M. Peimbert, G. Shields, J.
Shields, G. Steigman, E. Terlevich, and R. Terlevich for informative and
valuable discussions. The work of KAO is supported in  part by DOE grant
DE-FG02-94ER-40823. EDS is grateful for partial support from a NASA
LTSARP grant No. NAGW-3189.  

\clearpage
\begin{deluxetable}{lcccccc}
\scriptsize
\tablenum{1}
\tablewidth{0pt}
\tablecaption{Results From Synthetic Spectra \label{tbl-1}}
\tablehead{
\colhead{Variables} &
\colhead{Input} &
\colhead{3 Lines} &
\colhead{3 Lines} &
\colhead{5 Lines} &
\colhead{5 Lines} &
\colhead{6 Lines} \\
\colhead{} & \colhead{}     &
\colhead{$\tau$($\lambda$3889) $=$ 0}   &
\colhead{}&  
\colhead{abs(He~i) $=$ 0}   &
\colhead{}&  \colhead{}
}
\startdata
& & &Case 1 (B) \\
density  &10   &34.5$\pm$57.0 &33.9 $\pm$ 56.8  &7.1 $\pm$ 12.4   &5.8 $\pm$ 10.9   &5.0 $\pm$ 9.8 \\
$\tau$(3889)&0.0&(0.0)&0.018$\pm$0.039&0.003 $\pm$ 0.007&0.003 $\pm$ 0.007&0.003 $\pm$ 0.006 \\
abs(He~i) &0.0  &0.035$\pm$0.059  &0.037 $\pm$ 0.061&(0.0)  &0.047 $\pm$ 0.068&0.014 $\pm$ 0.021 \\
He/H     &0.080&0.0801$\pm$0.0018&0.0802$\pm$0.0018&0.0801$\pm$0.0009&0.0808$\pm$0.0013&0.0804$\pm$0.0009 \\
He/H (3) &0.080&0.0801$\pm$0.0018&0.0802$\pm$0.0018&0.0802$\pm$0.0011&0.0809$\pm$0.0014&0.0804$\pm$0.0011 \\        

& & &Case 2 (A) \\
density  &100  &91.9$\pm$92.7&90.9 $\pm$ 92.5  &69.4 $\pm$ 43.7  &59.7 $\pm$ 44.7  &58.4 $\pm$ 42.8 \\
$\tau$(3889)&0.0&(0.0)&0.028$\pm$0.047&0.022 $\pm$ 0.035&0.030 $\pm$ 0.041&0.029 $\pm$ 0.039 \\
abs(He~i) &0.0  &0.049$\pm$0.072&0.050 $\pm$ 0.074&(0.0)  &0.059 $\pm$ 0.076&0.021 $\pm$ 0.026 \\
He/H     &0.080&0.0811$\pm$0.025&0.0812$\pm$0.0026&0.0808$\pm$0.0014&0.0819$\pm$0.0019&0.0815$\pm$0.0015 \\
He/H (3) &0.080&0.0811$\pm$0.025&0.0812$\pm$0.0026&0.0808$\pm$0.0015&0.0819$\pm$0.0020&0.0813$\pm$0.0017 \\

& & &Case 3 (I) \\
density  &10   &35.2$\pm$57.7&34.7 $\pm$ 57.5  &33.7 $\pm$ 43.8  &26.7 $\pm$ 40.5  &22.0 $\pm$ 34.2 \\
$\tau$(3889)&0.1&(0.0)&0.020$\pm$0.044 &0.080 $\pm$ 0.058&0.090 $\pm$ 0.058&0.091 $\pm$ 0.055 \\
abs(He~i) &0.0  &0.036$\pm$0.060&0.038 $\pm$ 0.062&(0.0)  &0.038 $\pm$ 0.061&0.012 $\pm$ 0.020 \\
He/H     &0.080&0.0802$\pm$0.0018&0.0802$\pm$0.0019&0.0795$\pm$0.0014&0.0802$\pm$0.0016&0.0801$\pm$0.0012 \\
He/H (3) &0.080&0.0802$\pm$0.0018&0.0802$\pm$0.0019&0.0796$\pm$0.0015&0.0803$\pm$0.0018&0.0800$\pm$0.0015 \\

& & &Case 4 (J) \\
density  &10   &46.1$\pm$68.0&44.7 $\pm$ 67.4  & 9.0 $\pm$ 13.5  & 6.5 $\pm$ 11.5  & 7.1 $\pm$ 11.6 \\
$\tau$(3889)&0.0&(0.0)&0.030$\pm$0.052 &0.002 $\pm$ 0.005&0.003 $\pm$ 0.006&0.003 $\pm$ 0.006 \\
abs(He~i) &0.1  &0.094$\pm$0.093&0.097 $\pm$ 0.094&(0.0)  &0.114 $\pm$ 0.097&0.101 $\pm$ 0.035 \\
He/H     &0.080&0.0793$\pm$0.0024&0.0793$\pm$0.0019&0.0785$\pm$0.0009&0.0803$\pm$0.0017&0.0801$\pm$0.0012 \\
He/H (3) &0.080&0.0793$\pm$0.0024&0.0793$\pm$0.0019&0.0785$\pm$0.0011&0.0804$\pm$0.0018&0.0802$\pm$0.0013 \\

& & &Case 5 (H) \\
density  &10   &47.1$\pm$68.7&45.6 $\pm$ 68.1  &52.0 $\pm$ 50.5  &32.6 $\pm$ 43.9  &35.0 $\pm$ 43.6 \\
$\tau$(3889)&0.1&(0.0)&0.031$\pm$0.058 &0.059 $\pm$ 0.057&0.082 $\pm$ 0.059&0.079 $\pm$ 0.058 \\
abs(He~i) &0.1  &0.095$\pm$0.093&0.097 $\pm$ 0.095&(0.0)  &0.098 $\pm$ 0.092&0.095 $\pm$ 0.037 \\
He/H     &0.080&0.0793$\pm$0.0024&0.0793$\pm$0.0024&0.0775$\pm$0.0015&0.0795$\pm$0.0020&0.0794$\pm$0.0017 \\
He/H (3) &0.080&0.0793$\pm$0.0024&0.0793$\pm$0.0024&0.0776$\pm$0.0016&0.0796$\pm$0.0022&0.0795$\pm$0.0019 \\

& & &Case 6 (G) \\
density  &100  &111$\pm$103&108$\pm$104&135$\pm$75.5&101$\pm$75.2 &105$\pm$76.1 \\
$\tau$(3889)&0.1&(0.0)&0.041$\pm$0.052 &0.089 $\pm$ 0.096&0.140 $\pm$ 0.117&0.131 $\pm$ 0.115 \\
abs(He~i) &0.1  &0.111$\pm$0.104&0.115 $\pm$ 0.107&(0.0)  &0.116 $\pm$ 0.100&0.104 $\pm$ 0.041 \\
He/H     &0.080&0.0802$\pm$0.0031&0.0803$\pm$0.0032&0.0779$\pm$0.0019&0.0804$\pm$0.0028&0.0802$\pm$0.0025 \\
He/H (3) &0.080&0.0802$\pm$0.0031&0.0803$\pm$0.0032&0.0779$\pm$0.0020&0.0804$\pm$0.0029&0.0801$\pm$0.0026 \\

\enddata
\end{deluxetable}

\clearpage
\begin{deluxetable}{lcccccc}
\scriptsize
\tablenum{2}
\tablewidth{0pt}
\tablecaption{Results From Synthetic Spectra with He I $\lambda$3889 10\% Low \label{tbl-2}}
\tablehead{
\colhead{Variables} &
\colhead{Input} &
\colhead{3 Lines} &
\colhead{3 Lines} &
\colhead{5 Lines} &
\colhead{5 Lines} &
\colhead{6 Lines} \\
\colhead{} & \colhead{}     &
\colhead{$\tau$($\lambda$3889) $=$ 0}   &
\colhead{}&  
\colhead{abs(He~i) $=$ 0}   &
\colhead{}&  \colhead{}
}
\startdata
& & &Case 1 (B) \\
density  &10   &34.5$\pm$57.0 &33.9 $\pm$ 56.8  &1.3 $\pm$  2.8   &1.2 $\pm$  2.2   &0.1 $\pm$ 0.0 \\
$\tau$(3889)&0.0&(0.0)&0.018$\pm$0.039&0.024 $\pm$ 0.022&0.025 $\pm$ 0.022&0.020 $\pm$ 0.020 \\
abs(He~i) &0.0  &0.035$\pm$0.059  &0.037 $\pm$ 0.061&(0.0)  &0.032 $\pm$ 0.055&0.002 $\pm$ 0.006 \\
He/H     &0.080&0.0801$\pm$0.0018&0.0802$\pm$0.0018&0.0779$\pm$0.0008&0.0783$\pm$0.0011&0.0784$\pm$0.0007 \\
He/H (3) &0.080&0.0801$\pm$0.0018&0.0802$\pm$0.0018&0.0803$\pm$0.0010&0.0808$\pm$0.0012&0.0803$\pm$0.0010 \\        

& & &Case 2 (A) \\
density  &100  &91.9$\pm$92.7&90.9 $\pm$ 92.5  & 2.6 $\pm$ 10.0  & 1.1 $\pm$  4.1  & 1.1 $\pm$  5.8 \\
$\tau$(3889)&0.0&(0.0)&0.028$\pm$0.047&0.158 $\pm$ 0.055&0.163 $\pm$ 0.053&0.156 $\pm$ 0.052 \\
abs(He~i) &0.0  &0.049$\pm$0.072&0.050 $\pm$ 0.074&(0.0)  &0.053 $\pm$ 0.075&0.007 $\pm$ 0.014 \\
He/H     &0.080&0.0811$\pm$0.025&0.0812$\pm$0.0026&0.0800$\pm$0.0008&0.0809$\pm$0.0013&0.0804$\pm$0.0008 \\
He/H (3) &0.080&0.0811$\pm$0.025&0.0812$\pm$0.0026&0.0822$\pm$0.0010&0.0831$\pm$0.0015&0.0823$\pm$0.0011 \\

& & &Case 3 (I) \\
density  &10   &35.2$\pm$57.7&34.7 $\pm$ 57.5  & 1.2 $\pm$  3.1  & 0.7 $\pm$  1.5  & 1.4 $\pm$  3.9 \\
$\tau$(3889)&0.1&(0.0)&0.020$\pm$0.044 &0.221 $\pm$ 0.061&0.225 $\pm$ 0.063&0.208 $\pm$ 0.060 \\
abs(He~i) &0.0  &0.036$\pm$0.060&0.038 $\pm$ 0.062&(0.0)  &0.032 $\pm$ 0.055&0.003 $\pm$ 0.007 \\
He/H     &0.080&0.0802$\pm$0.0018&0.0802$\pm$0.0019&0.0779$\pm$0.0007&0.0784$\pm$0.0011&0.0783$\pm$0.0007 \\
He/H (3) &0.080&0.0802$\pm$0.0018&0.0802$\pm$0.0019&0.0803$\pm$0.0010&0.0808$\pm$0.0012&0.0803$\pm$0.0010 \\

& & &Case 4 (J) \\
density  &10   &46.1$\pm$68.0&44.7 $\pm$ 67.4  & 2.1 $\pm$  5.3  & 1.5 $\pm$  2.9  & 2.3 $\pm$  4.6 \\
$\tau$(3889)&0.0&(0.0)&0.030$\pm$0.052 &0.022 $\pm$ 0.021&0.023 $\pm$ 0.022&0.023 $\pm$ 0.021 \\
abs(He~i) &0.1  &0.094$\pm$0.093&0.097 $\pm$ 0.094&(0.0)  &0.084 $\pm$ 0.092&0.057 $\pm$ 0.035 \\
He/H     &0.080&0.0793$\pm$0.0024&0.0793$\pm$0.0019&0.0764$\pm$0.0008&0.0777$\pm$0.0015&0.0772$\pm$0.0011 \\
He/H (3) &0.080&0.0793$\pm$0.0024&0.0793$\pm$0.0019&0.0787$\pm$0.0010&0.0800$\pm$0.0017&0.0796$\pm$0.0012 \\

& & &Case 5 (H) \\
density  &10   &47.1$\pm$68.7&45.6 $\pm$ 68.1  & 1.4 $\pm$  5.2  & 1.0 $\pm$  2.8  & 2.7 $\pm$  6.6 \\
$\tau$(3889)&0.1&(0.0)&0.031$\pm$0.058 &0.220 $\pm$ 0.064&0.223 $\pm$ 0.063&0.219 $\pm$ 0.064 \\
abs(He~i) &0.1  &0.095$\pm$0.093&0.097 $\pm$ 0.095&(0.0)  &0.085 $\pm$ 0.091&0.058 $\pm$ 0.034 \\
He/H     &0.080&0.0793$\pm$0.0024&0.0793$\pm$0.0024&0.0764$\pm$0.0007&0.0777$\pm$0.0015&0.0772$\pm$0.0011 \\
He/H (3) &0.080&0.0793$\pm$0.0024&0.0793$\pm$0.0024&0.0787$\pm$0.0010&0.0800$\pm$0.0017&0.0796$\pm$0.0012 \\

& & &Case 6 (G) \\
density  &100  &111$\pm$103&108$\pm$104&2.1$\pm$ 7.9&1.7$\pm$ 6.2 &5.9$\pm$17.9 \\
$\tau$(3889)&0.1&(0.0)&0.041$\pm$0.052 &0.512 $\pm$ 0.096&0.503 $\pm$ 0.096&0.481 $\pm$ 0.133 \\
abs(He~i) &0.1  &0.111$\pm$0.104&0.115 $\pm$ 0.107&(0.0)  &0.129 $\pm$ 0.110&0.091 $\pm$ 0.036 \\
He/H     &0.080&0.0802$\pm$0.0031&0.0803$\pm$0.0032&0.0789$\pm$0.0008&0.0808$\pm$0.0018&0.0802$\pm$0.0011 \\
He/H (3) &0.080&0.0802$\pm$0.0031&0.0803$\pm$0.0032&0.0806$\pm$0.0010&0.0826$\pm$0.0019&0.0820$\pm$0.0013 \\

\enddata
\end{deluxetable}

\clearpage
\begin{deluxetable}{lcccccc}
\scriptsize
\tablenum{3}
\tablewidth{0pt}
\tablecaption{Results From Synthetic Spectra with He I $\lambda$7065 5\% Low \label{tbl-3}}
\tablehead{
\colhead{Variables} &
\colhead{Input} &
\colhead{3 Lines} &
\colhead{3 Lines} &
\colhead{5 Lines} &
\colhead{5 Lines} &
\colhead{6 Lines} \\
\colhead{} & \colhead{}     &
\colhead{$\tau$($\lambda$3889) $=$ 0}   &
\colhead{}&  
\colhead{abs(He~i) $=$ 0}   &
\colhead{}&  \colhead{}
}
\startdata
& & &Case 1 (B) \\
density  &10   &34.5$\pm$57.0 &33.9 $\pm$ 56.8  &0.2 $\pm$ 1.5    &0.3 $\pm$  1.6   &0.2 $\pm$ 1.1 \\
$\tau$(3889)&0.0&(0.0)&0.018$\pm$0.039&0.000 $\pm$ 0.001&0.000 $\pm$ 0.001&0.000 $\pm$ 0.018 \\
abs(He~i) &0.0  &0.035$\pm$0.059  &0.037 $\pm$ 0.061&(0.0)  &0.061 $\pm$ 0.081&0.010 $\pm$ 0.018 \\
He/H     &0.080&0.0801$\pm$0.0018&0.0802$\pm$0.0018&0.0796$\pm$0.0007&0.0805$\pm$0.0014&0.0799$\pm$0.0008 \\
He/H (3) &0.080&0.0801$\pm$0.0018&0.0802$\pm$0.0018&0.0803$\pm$0.0010&0.0813$\pm$0.0015&0.0805$\pm$0.0010 \\        

& & &Case 2 (A) \\
density  &100  &91.9$\pm$92.7&90.9 $\pm$ 92.5  &36.3 $\pm$ 28.1  &30.9 $\pm$ 26.5  &30.8 $\pm$ 25.5 \\
$\tau$(3889)&0.0&(0.0)&0.028$\pm$0.047&0.006 $\pm$ 0.013&0.008 $\pm$ 0.015&0.007 $\pm$ 0.014 \\
abs(He~i) &0.0  &0.049$\pm$0.072&0.050 $\pm$ 0.074&(0.0)  &0.073 $\pm$ 0.084&0.026 $\pm$ 0.028 \\
He/H     &0.080&0.0811$\pm$0.025&0.0812$\pm$0.0026&0.0815$\pm$0.0012&0.0827$\pm$0.0017&0.0821$\pm$0.0013 \\
He/H (3) &0.080&0.0811$\pm$0.025&0.0812$\pm$0.0026&0.0815$\pm$0.0013&0.0827$\pm$0.0018&0.0820$\pm$0.0015 \\

& & &Case 3 (I) \\
density  &10   &35.2$\pm$57.7&34.7 $\pm$ 57.5  &23.4 $\pm$ 31.0  &18.0 $\pm$ 27.7  &15.9 $\pm$ 25.1 \\
$\tau$(3889)&0.1&(0.0)&0.020$\pm$0.044 &0.028 $\pm$ 0.029&0.032 $\pm$ 0.029&0.032 $\pm$ 0.028 \\
abs(He~i) &0.0  &0.036$\pm$0.060&0.038 $\pm$ 0.062&(0.0)  &0.041 $\pm$ 0.062&0.013 $\pm$ 0.020 \\
He/H     &0.080&0.0802$\pm$0.0018&0.0802$\pm$0.0019&0.0796$\pm$0.0012&0.0804$\pm$0.0014&0.0802$\pm$0.0011 \\
He/H (3) &0.080&0.0802$\pm$0.0018&0.0802$\pm$0.0019&0.0798$\pm$0.0013&0.0806$\pm$0.0016&0.0802$\pm$0.0013 \\

& & &Case 4 (J) \\
density  &10   &46.1$\pm$68.0&44.7 $\pm$ 67.4  & 0.3 $\pm$ 1.6   & 0.4 $\pm$ 1.6   & 0.5 $\pm$  1.7 \\
$\tau$(3889)&0.0&(0.0)&0.030$\pm$0.052 &0.000 $\pm$ 0.000&0.000 $\pm$ 0.001&0.000 $\pm$ 0.001 \\
abs(He~i) &0.1  &0.094$\pm$0.093&0.097 $\pm$ 0.094&(0.0)  &0.138 $\pm$ 0.112&0.094 $\pm$ 0.039 \\
He/H     &0.080&0.0793$\pm$0.0024&0.0793$\pm$0.0019&0.0781$\pm$0.0007&0.0802$\pm$0.0019&0.0796$\pm$0.0011 \\
He/H (3) &0.080&0.0793$\pm$0.0024&0.0793$\pm$0.0019&0.0787$\pm$0.0010&0.0809$\pm$0.0020&0.0802$\pm$0.0013 \\

& & &Case 5 (H) \\
density  &10   &47.1$\pm$68.7&45.6 $\pm$ 68.1  &35.5 $\pm$ 35.2  &21.7 $\pm$ 29.4  &23.3 $\pm$ 29.3 \\
$\tau$(3889)&0.1&(0.0)&0.031$\pm$0.058 &0.021 $\pm$ 0.027&0.030 $\pm$ 0.029&0.029 $\pm$ 0.029 \\
abs(He~i) &0.1  &0.095$\pm$0.093&0.097 $\pm$ 0.095&(0.0)  &0.103 $\pm$ 0.095&0.096 $\pm$ 0.038 \\
He/H     &0.080&0.0793$\pm$0.0024&0.0793$\pm$0.0024&0.0778$\pm$0.0012&0.0797$\pm$0.0018&0.0796$\pm$0.0015 \\
He/H (3) &0.080&0.0793$\pm$0.0024&0.0793$\pm$0.0024&0.0780$\pm$0.0014&0.0799$\pm$0.0020&0.0798$\pm$0.0017 \\

& & &Case 6 (G) \\
density  &100  &111$\pm$103&108$\pm$104&115$\pm$64.7&81.4$\pm$65.3 &85.3$\pm$65.4 \\
$\tau$(3889)&0.1&(0.0)&0.041$\pm$0.052 &0.048 $\pm$ 0.067&0.082 $\pm$ 0.082&0.076 $\pm$ 0.079 \\
abs(He~i) &0.1  &0.111$\pm$0.104&0.115 $\pm$ 0.107&(0.0)  &0.124 $\pm$ 0.106&0.106 $\pm$ 0.042 \\
He/H     &0.080&0.0802$\pm$0.0031&0.0803$\pm$0.0032&0.0782$\pm$0.0018&0.0809$\pm$0.0027&0.0805$\pm$0.0023 \\
He/H (3) &0.080&0.0802$\pm$0.0031&0.0803$\pm$0.0032&0.0783$\pm$0.0019&0.0809$\pm$0.0028&0.0806$\pm$0.0024 \\

\enddata
\end{deluxetable}
 
\clearpage

\appendix

\section{Monte Carlo Estimates of Reddening and Underlying Balmer Absorption}

Here we would like to describe the Monte Carlo procedure we use 
for determining the corrections for reddening and underlying 
stellar absorption in the Hydrogen lines.  Beginning with an 
observed line flux $F(\lambda)$, and an equivalent width 
$W(\lambda)$, we can parameterize the correction for underlying stellar 
absorption as
\beq
X_A(\lambda) = F(\lambda) ({W(\lambda) + a_{HI}) \over W(\lambda)})
\eeq
The parameter $a_{HI}$ is expected to be relatively insensitive to 
wavelength because all of the Balmer lines should be saturated
in the stars which are producing the underlying continuum.
As described in section 2, the reddening correction is applied 
to determine 
the intrinsic line intensity $I(\lambda)$ relative to $H\beta$
\beq
X_R(\lambda) = {I(\lambda) \over I(H\beta)} = {X_A(\lambda) \over X_A (H\beta)}
10^{f(\lambda) C(H\beta)}
\eeq
We assume the intrinsic Balmer
line ratios calculated by Hummer \& Storey (1987), and 
we use the reddening function, $f(\lambda)$,
normalized at H$\beta$, from the Galactic reddening law of Seaton (1979),
as parameterized by Howarth (1983), assuming a value of
R $\equiv$ A$_{\rm V}$/E$_{\rm B-V}$ = 3.2.
By comparing $X_R(\lambda)$ to theoretical values, $X_T(\lambda)$, we 
determine the parameters $a_{HI}$ and $C(H\beta)$ self consistently, and
run  a Monte Carlo over the input data to test the robustness of the
solution  and to determine the systematic uncertainty associated with
these  corrections. 

For definiteness, we list here the theoretical ratios we use:
\begin{eqnarray}
X_T(6563) & = & ~~0.3862(\log T_4)^2~ -~0.4817 \log T_4 ~+~ 2.86
\nonumber
\\ 
X_T(4340) & = & -0.01655(\log T_4)^2 -0.02824 \log T_4 + 0.468
\nonumber
\\
X_T(4101) & = & -0.01655(\log T_4)^2 -0.02159 \log T_4 + 0.259
\end{eqnarray}
where the temperature is $T_4 = T/10^4$K. The values for $f(\lambda)$ we
take are 
\beq
f(6563) = -0.329 \qquad f(4340) = 0.137 \qquad f(4101) = 0.193
\eeq

We begin therefore with four input fluxes $F(\lambda)$ along with their 
associated observational (statistical) uncertainties and four equivalent 
widths (and their observational uncertainties).  In addition, the theoretical 
ratios are temperature dependent, so we must add the temperature and its 
uncertainty as an additional observational input.  From these, the ratios, 
$X_R(\lambda)$ for $H\alpha, H\gamma$, and $H\delta$ are obtained. 
The $\chi^2$ statistic is defined by
\beq
\chi^2 = \sum_\lambda {\bigl(X_R^2(\lambda) - X_T^2(\lambda)\bigr) \over
\sigma_{X_R}^2(\lambda)} 
\eeq
where $\sigma_{X_R}(\lambda)$ is the derived uncertainty in $X_R$. 
The minimization of $\chi^2$, allows us to determine the values of
$a_{HI}$ and $C(H\beta)$. The uncertainties in the two outputs are
determined by varying the solution so that $\chi^2(a\pm\sigma_a) -
\chi^2(a) = 1$.
$\sigma_{C(H\beta)}$ is similarly determined.

As we indicated above, we further test this solution and its robustness 
by running a Monte Carlo on the input data. This also allows us to better 
determine the systematic uncertainty in the output parameters $a_{HI}$
and  $C(H\beta)$. The data for the Monte Carlo are generated from the
input data, $F(\lambda), W(\lambda)$ and temperature and the uncertainties
in these quantities. A Gaussian distribution of input  values centered on
the observed values with a spread determined by their  observational
uncertainties. A new set of data is then  randomly
generated by picking input values from these Gaussian distributions.
Consequently, after running the $\chi^2$ minimization  procedure, new
values for the output parameters $a_{HI}$ and $C(H\beta)$ are found.  Our
Monte Carlo produces 1000 randomly generated data sets from which  we can
produce a distribution of solutions for $a_{HI}$ and $C(H\beta)$. One 
would expect that the mean of the solutions for $a_{HI}$ and $C(H\beta)$
tracks the original solution based on the actual observational data.   The
spread in these allows us to test the systematic uncertainty  associated
with these quantities. 

\clearpage

\section{Monte Carlo Estimates of Reddening and Underlying Balmer
Absorption in Real Observations}

Here we provide examples of deriving reddening and underlying
stellar absorption from emission line spectra.  We take as a
examples, the observations of SBS1159$+$545, SBS1415$+$437,
and SBS1420$+$544 as reported in IT98.  These three spectra are
all reported with very high accuracy; the brightest lines
are reported with errors of less than one percent.  Since the
emission line equivalent widths for all of the Balmer lines
are required as input and since only the emission line equivalent
width for H$\beta$ is reported, we have had to estimate these
from the relative line strengths and the spectra shown in
figures.  The fractional uncertainties in the equivalent widths
were assumed to be twice as large as the fractional uncertainties
reported in the relative intensities.

We used the values reported in
IT98 and calculated C(H$\beta$) and $a_{HI}$ for the three
targets.  Our results are shown in Figure B-1.  Here we display
the originally observed Balmer line ratios, the corrected ratios
reported in IT98 and our own solutions.  There are two important
points to note.  First note that the IT98 corrected values for the
H$\gamma$/H$\beta$ and H$\delta$/H$\beta$ ratios are several $\sigma$
away from the theoretical values.  In the case of SBS 1159+545,
this is because these ratios were already higher than the theoretical
ratios, and correcting for reddening and underlying absorption only
increases the ratios.  In the other two cases, the original ratios
were very close to the theoretical ratios, but correcting the
H$\alpha$/H$\beta$ ratio for reddening has caused these ratios to
exceed their theoretical ratios.  Since the deviations from the
theoretical line ratios are large (yielding generally high values of
the $\chi ^2$), we conclude that the uncertainties
in the reported emission lines are underestimated.

The second point to note in Figure B-1 is the difference between our
solutions are those of IT98.  In all three cases, the IT98 solution
yields very good agreement with theory in the H$\alpha$/H$\beta$
ratio (better than ours) but not as good in the other lines.  It would
appear that the weighting scheme used by IT98 favors this line ratio
more than would be called for by the relative errors in the line
ratios.

\clearpage

\section{Monte Carlo Estimates of Self-Consistent Helium Line Ratios}

For \he4, we follow an analogous procedure to that described above.
We again start with a set of observed quantities: line intensities 
$I(\lambda)$ which include the reddening correction previously 
determined and its associated uncertainty which also includes the 
uncertainty in $C(H\beta)$; the equivalent width $W(\lambda)$; and 
temperature $t$. The Helium line intensities are scaled to $H\beta$ 
and the singly ionized helium abundance is given by 
\beq
y^+(\lambda) = {I(\lambda) \over I(H\beta)} {E(H\beta) \over
E(\lambda)} ({W(\lambda) + a_{He~i})
\over W(\lambda)}) {1\over (1+\gamma)} {1\over f(\tau)}
\label{y+}
\eeq
where $E(\lambda)/E(H\beta)$ is the theoretical emissivity scaled 
to $H\beta$.  The expression (\ref{y+}), also contains a correction 
factor for underlying stellar absorption, parameterized now by $a_{HeI}$, 
a density dependent collisional correction factor, 
$(1+\gamma)^{-1}$, and a florescence correction which 
depends on the optical depth $\tau$.  Thus $y^+$ implicitly 
depends on three unknowns, the electron density, $n$, $a_{HeI}$, and
$\tau$.
 
To be definite, we list here the necessary components in expression
(\ref{y+}). The theoretical emissivities scaled 
to $H\beta$ are taken from Smits (1996):
\begin{eqnarray}
E(H\beta)/E(3889) & = & 0.9072 T^{-0.1715} \nonumber \\
E(H\beta)/E(4026) & = & 4.3166 T^{0.0847} \nonumber \\
E(H\beta)/E(4471) & = & 2.0094 T^{0.1259} \nonumber \\
E(H\beta)/E(5876) & = & 0.7355 T^{0.2298} \nonumber \\
E(H\beta)/E(6678) & = & 2.5861 T^{0.2475} \nonumber \\
E(H\beta)/E(7065) & = & 4.3588 T^{-0.3456} 
\end{eqnarray}
 Our expressions for the
collisional correction
$\gamma$, are taken from Kingdon \& Ferland (1995).  We list them here
for completeness. They are:
\begin{eqnarray}
\gamma(3889)&  = &  (9.34 T_4^{-0.92}e^{-3.699/T_4} + 1.64
T_4^{-0.79}e^{-4.379/T_4} + 0.83 T_4^{-0.40}e^{-4.545/T_4} \nonumber \\
& & + 0.51 T_4^{-1.05}e^{-4.818/T_4} + 0.39 T_4^{-0.36}e^{-4.900/T_4})/D
\nonumber \\
\gamma(4026) &  = & (6.92 T_4^{-0.45}e^{-4.900/T_4} )/D \nonumber \\
\gamma(4471)&  = &  (6.95 T_4^{0.15}e^{-4.545/T_4} + 0.22
T_4^{-0.55}e^{-4.884/T_4} + 0.98 T_4^{-0.45}e^{-4.901/T_4})/D \nonumber \\
\gamma(5876)&  = &  (6.78 T_4^{0.07}e^{-3.776/T_4} + 1.67
T_4^{-0.15}e^{-4.545/T_4} + 0.60 T_4^{-0.34}e^{-4.901/T_4})/D \nonumber \\
\gamma(6678)&  = &  (3.15 T_4^{-0.54}e^{-3.776/T_4} + 0.51
T_4^{-0.51}e^{-4.545/T_4} + 0.20 T_4^{-0.66}e^{-4.901/T_4})/D \nonumber \\
\gamma(7065)&  = &  (38.09 T_4^{-1.09}e^{-3.364/T_4} + 2.80
T_4^{-1.06}e^{-3.699/T_4})/D
\end{eqnarray}
where $D = 1 + 3130 n^{-1} T_4^{-0.50}$.
The corrections for florescence are given in terms of the optical depth
for the He I $\lambda \lambda 3389$ line. We use the IT98 fit of the
Robbins (1968) enhancement factors:
\begin{eqnarray}
f(3889) & = & 2.25 \times 10^{-7} \tau^4 - 3.87 \times 10^{-5} \tau^3 +
2.39 \times 10^{-3} \tau^2 -0.069 \tau + 1 \nonumber \\
f(4026) & = & 1 \nonumber \\
f(4471) & = & 1 + 0.001 \tau \nonumber \\
f(5876) & = & 1 + 0.0049 \tau \nonumber \\
f(6678) & = & 1  \nonumber \\
f(7065) & = & 1 + 0.4 \tau^{0.55}
\end{eqnarray}
$f(4026)$ is not given by IT98, but is  assumed to be 1 because it is a
singlet line (as is the case for $\lambda$6678). 

Once the individual values for $y^+(\lambda)$ are determined, we can
begin the process for self-consistently determining the physical
parameters.  As described in the text, we may wish to consider 3,5, or 6
different \he4  emission lines. Depending on the number of lines used, we
next determine  the average helium abundance.
$\bar y$, 
\beq
\bar y = \sum_{\lambda} {y^+(\lambda) \over \sigma(\lambda)^2} / \sum_{\lambda}
{1 \over \sigma(\lambda)^2}
\eeq
This is a weighted average, where the uncertainty
$\sigma(\lambda)$ is found by propagating the uncertainties in the
observational  quantities stemming from the observed line fluxes (which
already contains  the uncertainty due to C(H$\beta$), the equivalent
widths, and input temperature. Since the average, $\bar y$, depends on
the parameters, $n,
\tau$ and
$a_{HeI}$, we must make an initial estimate for these.

{}From ${\bar y}$, we can define a $\chi^2$ as the deviation of the
individual He abundances $y^+(\lambda)$ from the average,
\beq
\chi^2 = \sum_{\lambda} {(y^+(\lambda) - {\bar y})^2 \over
\sigma(\lambda)^2}
\label{chi4}
\eeq
We then minimize $\chi^2$, to determine $n, a_{HeI}$, and $\tau$. 
Uncertainties in the output parameters are determined as in the
case for $a_{HI}$ and $C(H\beta)$, that is by varying the outputs until 
$\Delta \chi^2= 1$. Propagation in the latter uncertainties give us a
reasonable handle on the systematic uncertainties in our final result for
$y^+$.

This procedure differs somewhat from that proposed by IT98, in that the 
$\chi^2$ above (\ref{chi4}) is a straight weighted average, whereas IT98 
minimize the differences between ratios of He abundances from pairs of
He~I lines (referenced to one wavelength, 
typically 4471).  When the reference line is  particularly sensitive to a
systematic effect such as underlying  stellar absorption, the uncertainty
propagates to all lines this way. In our case, the individual
uncertainties in the line strengths  are kept separate.

Finally, as in the case for the hydrogen lines, we have performed a 
Monte-Carlo simulation of the data to test the robustness of the 
solution for $n, a_{HeI}$, and $\tau$ from the $\chi^2$ minimization and
the  true uncertainty in these quantities.
As before, starting with the observational inputs and their stated
uncertainties, we have generated a data set which is Gaussian
distributed  for the 6 observed He
emission lines  (plus the temperature). From each distribution, we
randomly select  a set of input values and run the
$\chi^2$ minimization. The selection  of data is repeated 1000 times. We
thus obtain a distribution of  solutions for $n, a_{HeI}$, and $\tau$,
and we compare the mean and dispersion  of these distributions with the
initial solution for these quantities.

\newpage
\noindent{\bf{Figure Captions}}

\vskip.3truein

\begin{itemize}

\item[]
\begin{enumerate}
\item[]
\begin{enumerate}

\item[{\bf Figure 1:}] A comparison of the observed and corrected
hydrogen Balmer emission line ratios for three synthetic cases
(with no errors).  The open circles show the deviations of
the original synthesized spectra from the theoretical ratios 
in terms of the synthesized uncertainties (assumed to be 2\%
for all lines).
The filled circles show the corrected values in the same manner.
Note that the corrections for reddening and underlying absorption
all have the same sense for all three line ratios, i.e., the
H$\alpha$/H$\beta$ line ratio increases for increased reddening
and underlying absorption and the higher order Balmer line
ratios all decrease for both effects.  This covariance results
in decreased diagnostic power as shown in Figure 2. 

\item[{\bf Figure 2:}] The results of Monte Carlo simulations
of solutions for the reddening and underlying absorption from 
hydrogen Balmer emission line ratios for three synthetic cases.
Each small point is the solution derived from a different
realization of the same input spectrum (with 2\% errors in
both emission line flux and equivalent width). The original
input spectra had reddening with C(H$\beta$) $=$ 0.1 and 
$a_{HI}$ = 2 \AA . The large filled 
point with error bars shows the mean result with 1$\sigma$ errors
derived from the dispersion in the solutions.  
Note that the covariance of the two parameters leads to error
ellipses. The Monte
Carlo simulations find the correct solutions, but the error
bars appropriate to these solutions are about twice the
size of the errors inferred from single $\chi^2$ minimizations.
Note that the covariance of the two parameters leads to error
ellipses.  Note also that, given the input assumptions, that
the constraints on the underlying absorption are stronger 
in absolute terms for the lower emission line equivalent width 
cases.

\item[{\bf Figure 3:}] Graphs showing the IT98 fits to the
florescence enhancement figures reported in Robbins (1968).
The important point to note is that regime of the calculations
is mostly at far larger values of $\tau$(3889) than is relevant
for work with giant extragalactic HII regions.  More detailed
work concentrating on the relevant regime is needed.

\item[{\bf Figure 4:}] Histograms showing the effects on helium
emission lines due to changes in physical parameters.  The baseline
model is a photoionized gas at 15,000 K with a density of 10 cm$^{-3}$,
a negligible optical depth in the $\lambda$3889 line, and no 
underlying absorption in the helium lines (the underlying spectrum
has a $\lambda$$^{-2}$ slope and the equivalent width of the H$\beta$
emission line is 100 \AA ).  The top panel shows that
the helium lines are very insensitive to variations in temperature,
with differences of order 1\% or less for a 500 K variations.
The second panel shows that all of the helium lines are sensitive to 
an increase in density, but that the $\lambda$7065 line is far more
sensitive than the other lines.  The third panel shows that only the
$\lambda$3889 and $\lambda$7065 lines have significant sensitivity
to optical depth effects.  The bottom panel shows that all lines 
are very sensitive to small increases in the underlying absorption.

\item[{\bf Figure 5:}] Results of modeling of 6 synthetic He~I
line observations.  The 
four panels show the results of a density = 10,
$a_{HeI}$ = 0, and $\tau$(3889) $=$ 0 model. 
The solid lines show the input values (e.g., He/H = 0.080)
for the original calculated spectrum.  The solid circles 
(with error bars) show
the results of the $\chi ^2$ minimization solution 
(with calculated  errors) for
the original synthetic input spectrum.
The small points show the results of Monte Carlo realizations
of the original input spectrum.  
The solid squares (with error bars) show the means and dispersions
of the output values for the $\chi ^2$ minimization solutions of
the Monte Carlo realizations.

\item[{\bf Figure 6:}] Similar plot to Figure 5 except that the
density $=$ 100 as opposed to 10 as in Figure 5.
 
\item[{\bf Figure 7:}] Similar plot to Figure 6 except that the
underlying absorption is 0.1 \AA and $\tau$(3889) $=$ 0.1.

\item[{\bf Figure 8:}] Similar plot to Figure 6 except that 
I($\lambda$3889) and EW($\lambda$3889) are artificially decreased 
by 10\%.
 
\item[{\bf Figure 9:}] Similar plot to Figure 6 except that 
I($\lambda$7065) and EW($\lambda$7065) are artificially decreased 
by 5\%.
 
\item[{\bf Figure 10:}] Similar plot to Figure 7 except that 
I($\lambda$7065) and EW($\lambda$7065) are artificially decreased 
by 5\%.

\item[{\bf Figure B-1:}] A comparison of the observed and corrected
hydrogen Balmer emission line ratios for three blue compact galaxies
from the sample of IT98.  The open circles show the deviations of
the original observations from the theoretical ratios in terms of the
uncertainties in the final reported corrected values
(the H$\alpha$/H$\beta$ ratios for SBS 1415+437 and SBS 1420+544 are
off scale at $+$ 33 $\sigma$ and $+$ 25 $\sigma$ respectively).
The filled circles show the corrected values in the same manner.
Note that in all three cases the corrected H$\gamma$/H$\beta$ and
H$\delta$/H$\beta$ ratios are several $\sigma$ away from the theoretical
values.  In the case of SBS 1159+545, this is because these ratios
were already higher than the theoretical ratios, and correcting for
reddening and underlying absorption only increases the ratios.
In the other two cases, the original ratios were very close to the
theoretical ratios, but correcting the H$\alpha$/H$\beta$ ratio
for reddening has caused these ratios to exceed their theoretical
ratios.  Since the deviations from the theoretical line ratios are
large, we conclude that the uncertainties in the reported emission
lines are underestimated.

\end{enumerate}
\end{enumerate}
\end{itemize}

\newpage

\begin{figure}
\vskip -2cm
\hskip -2cm
\epsfig{file=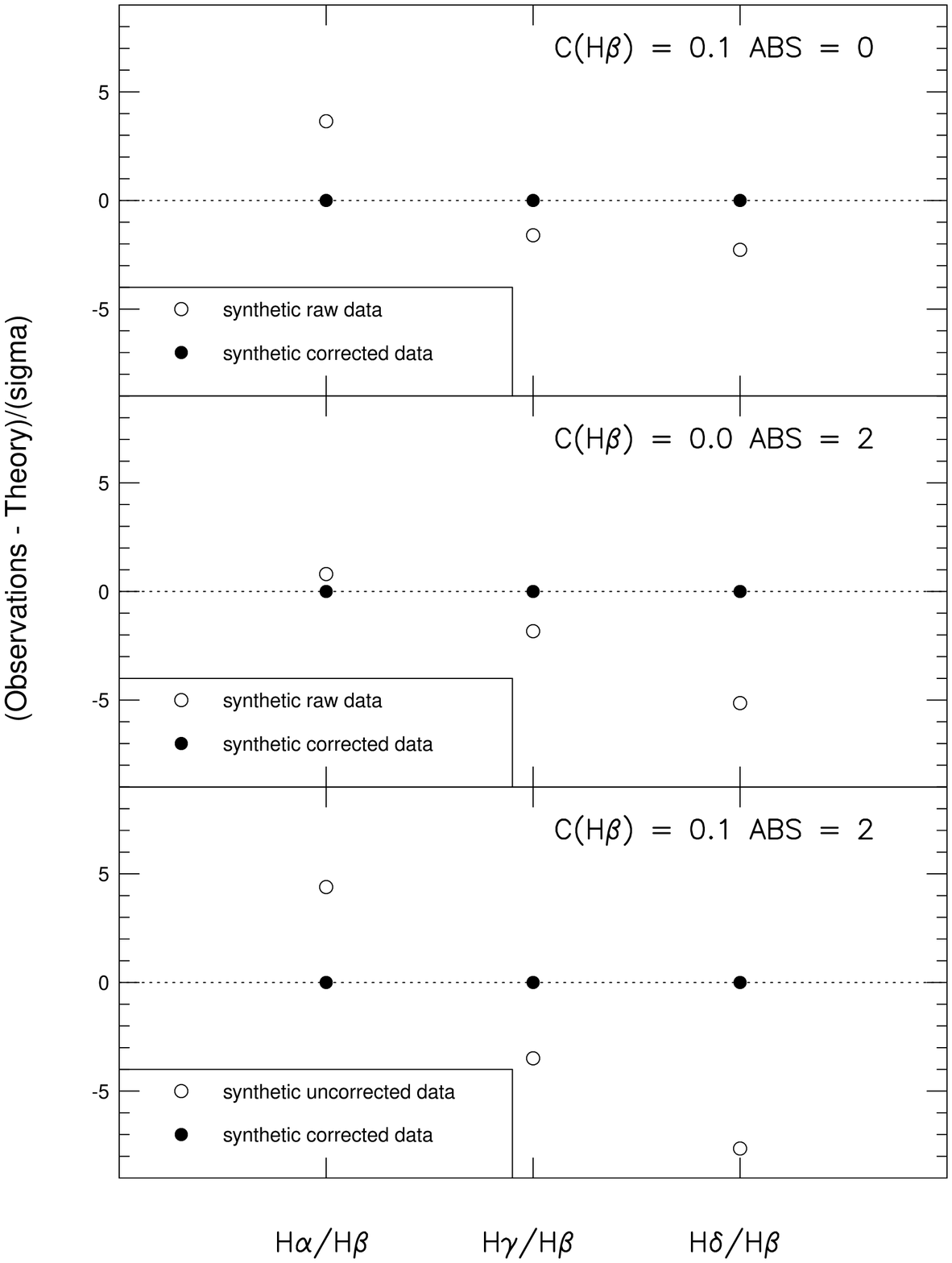}
\end{figure}

\newpage

\begin{figure}
\vskip -2cm
\hskip -2cm
\epsfig{file=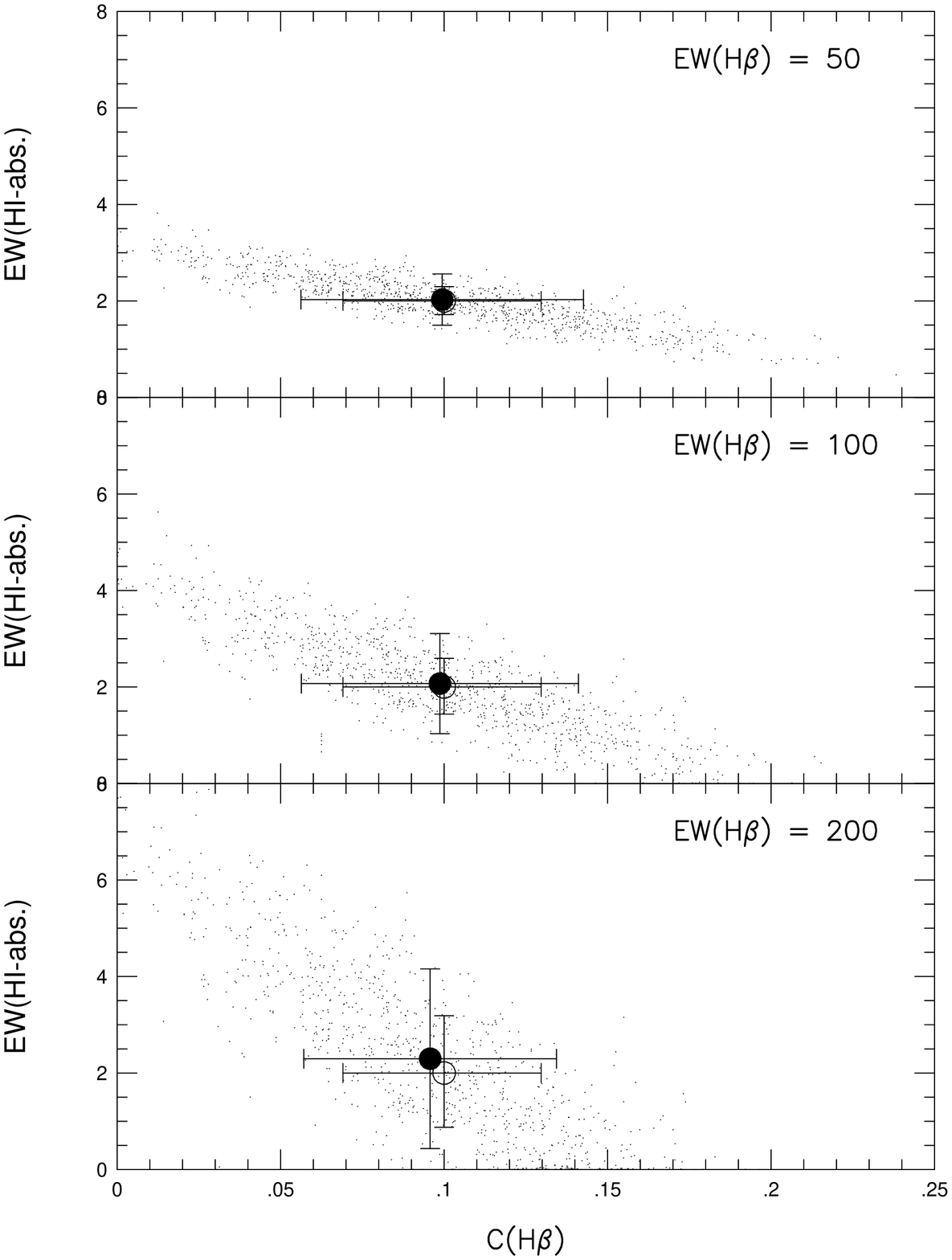}
\end{figure}

\newpage

\begin{figure}
\vskip -2cm
\hskip -2cm
\epsfig{file=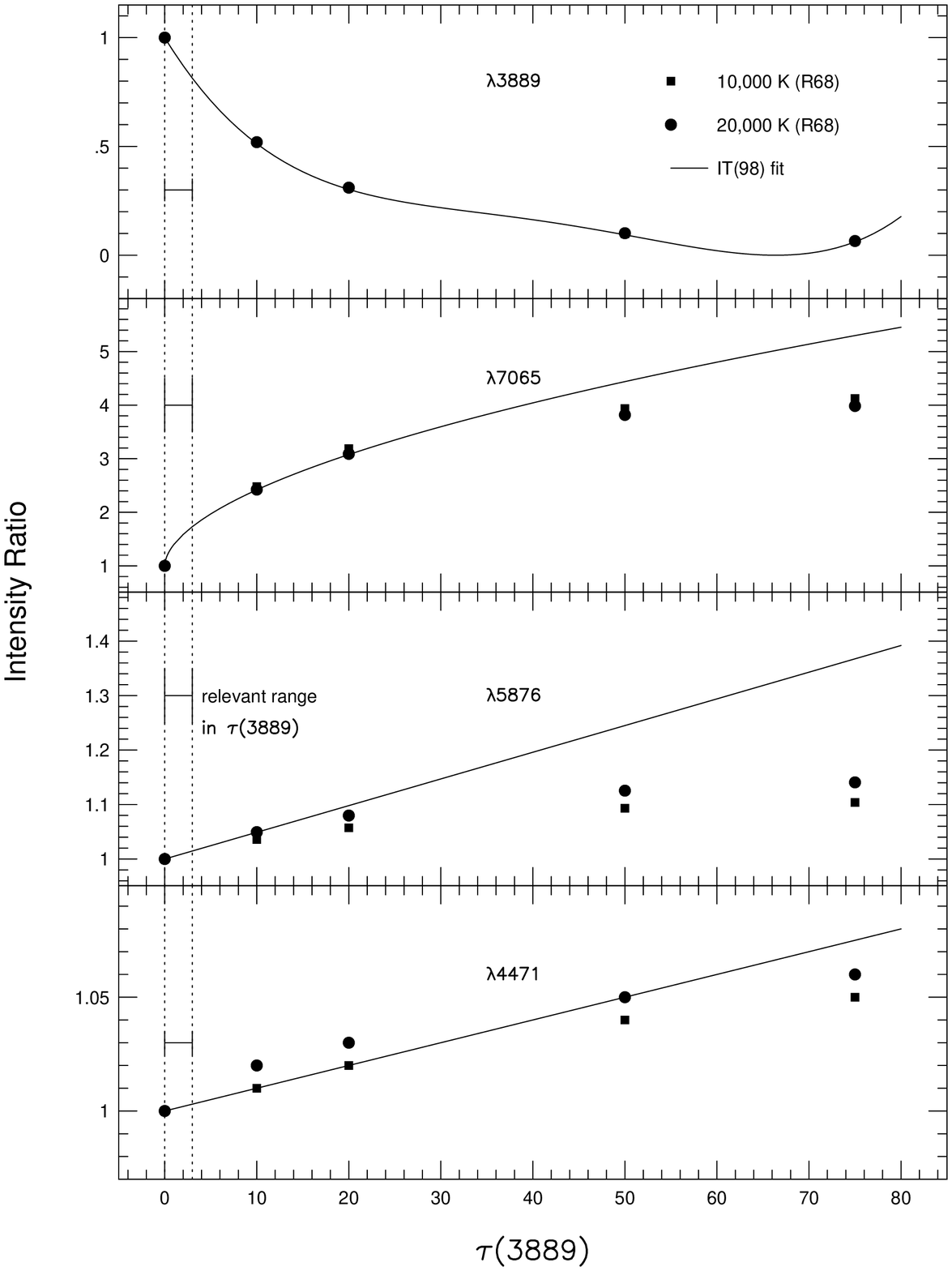}
\end{figure}

\newpage

\begin{figure}
\vskip -2cm
\hskip -2cm
\epsfig{file=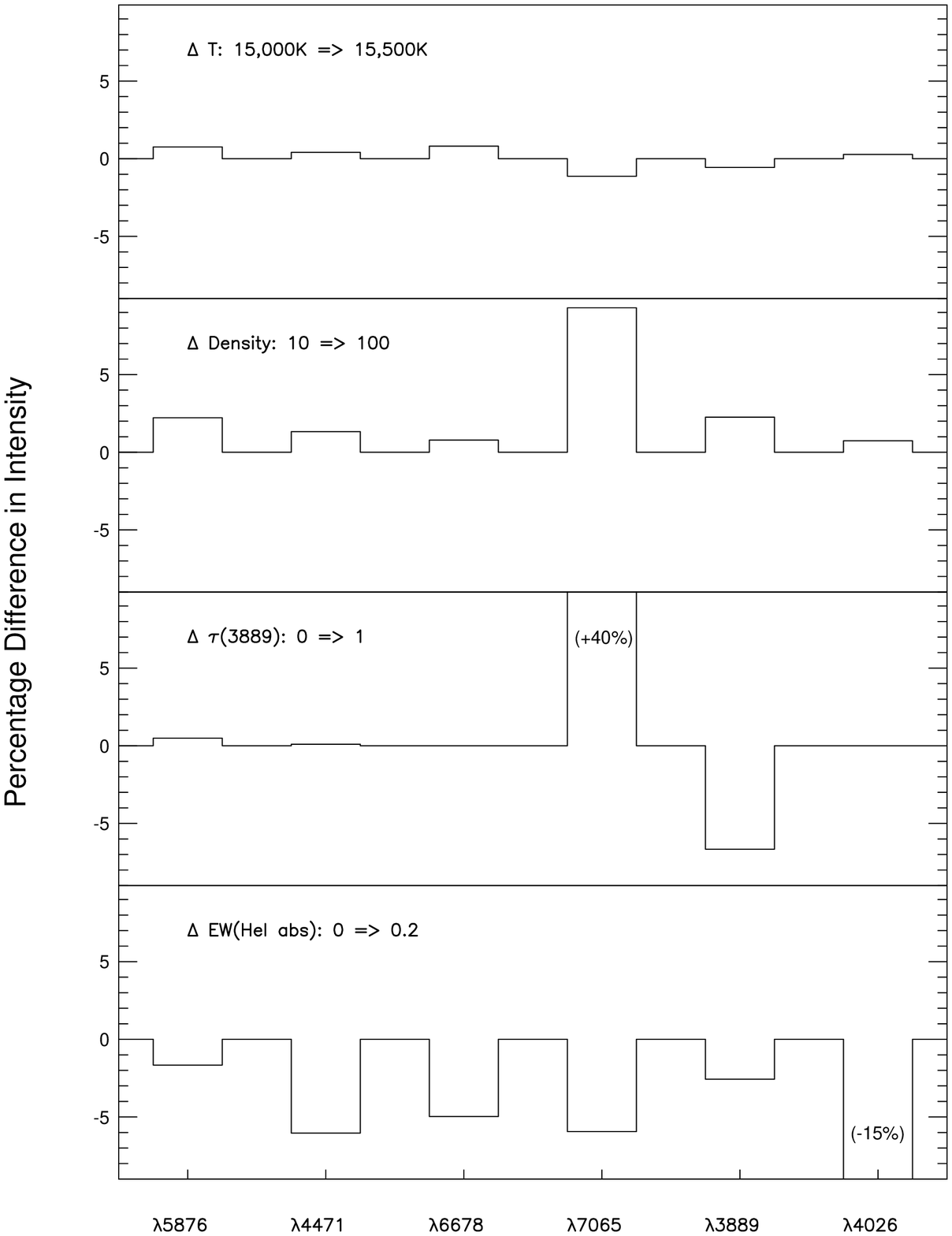}
\end{figure}

\newpage

\begin{figure}
\vskip -2cm
\hskip -2cm
\epsfig{file=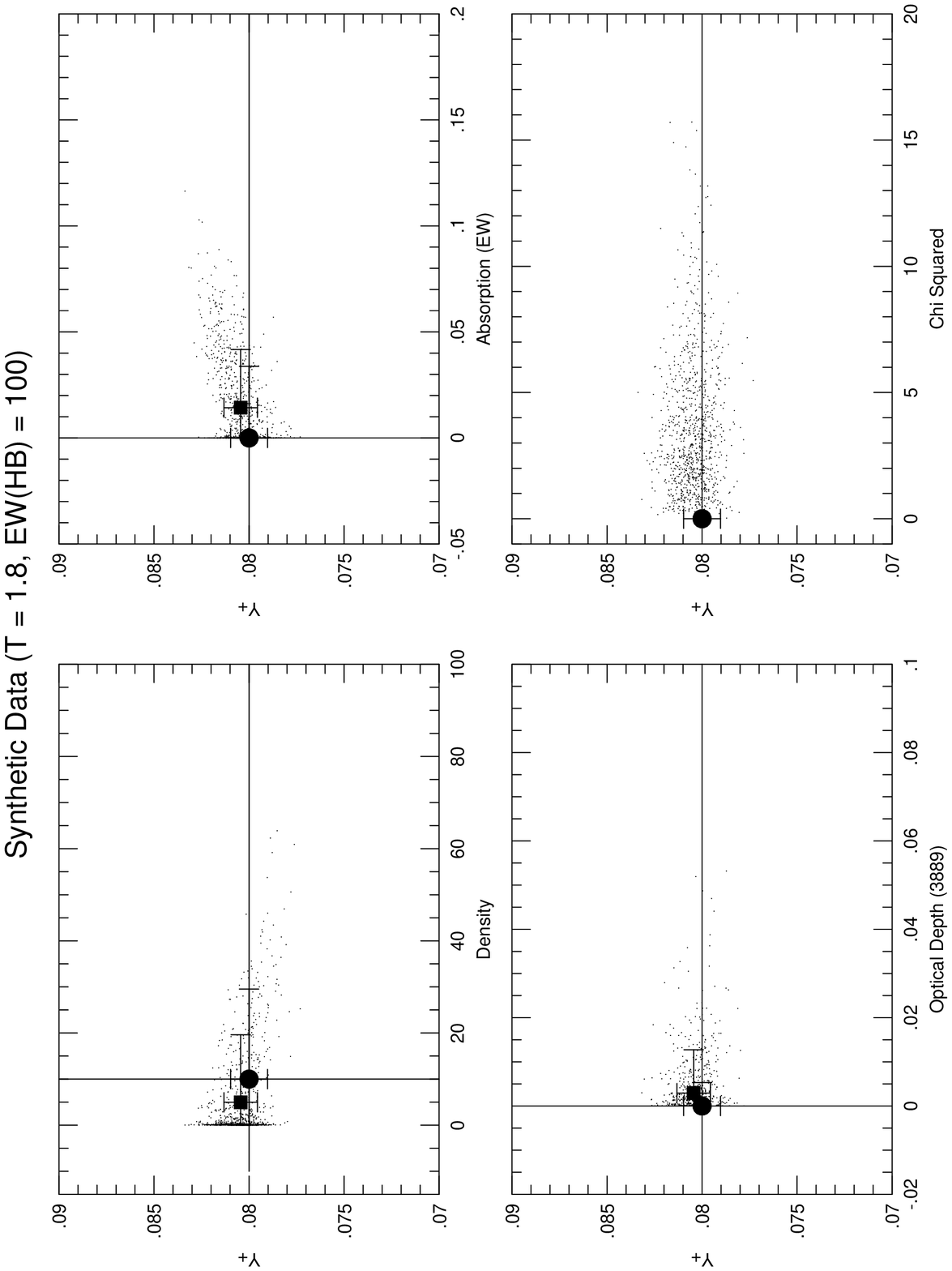}
\end{figure}

\newpage

\begin{figure}
\vskip -2cm
\hskip -2cm
\epsfig{file=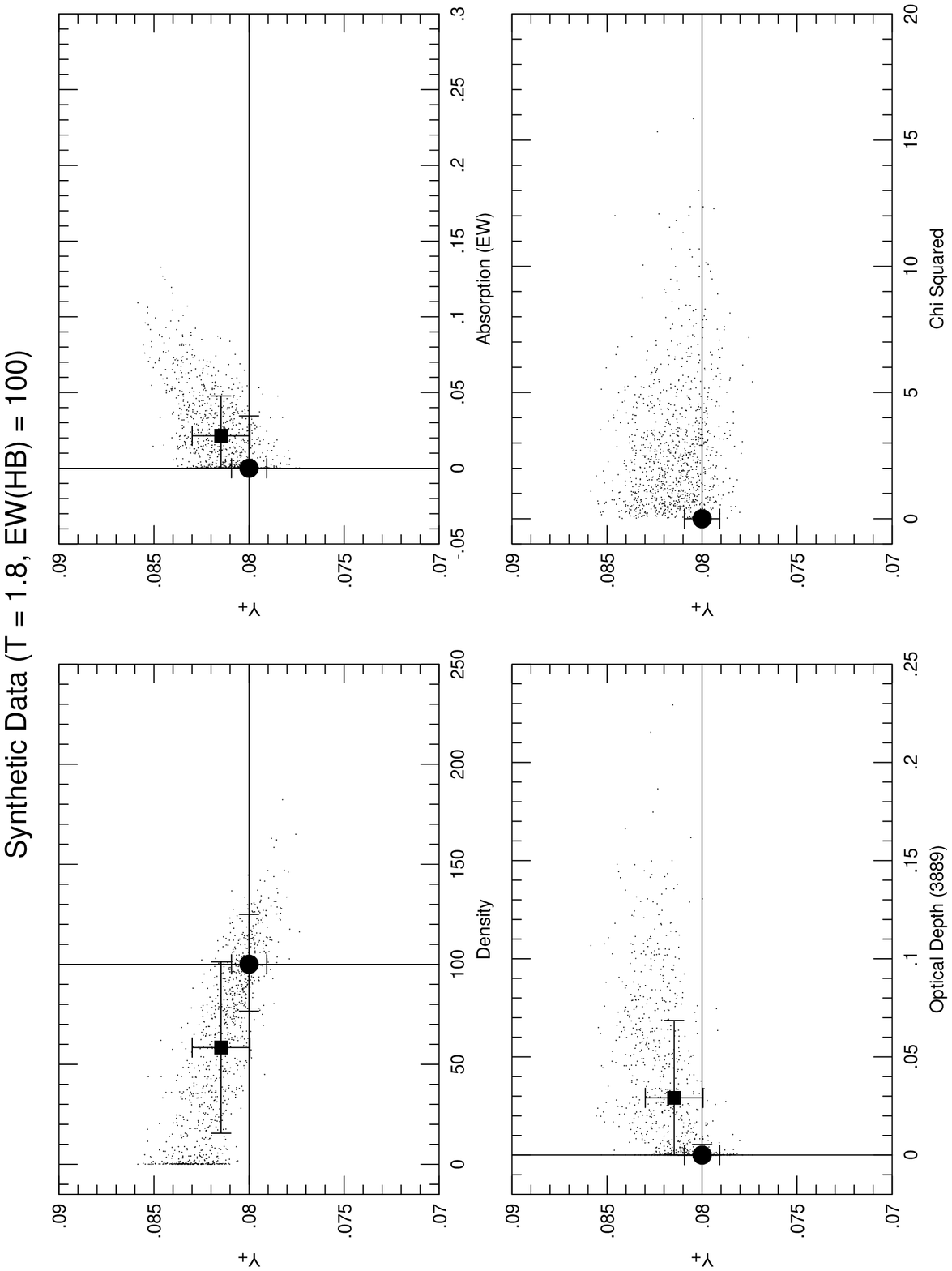}
\end{figure}

\newpage

\begin{figure}
\vskip -2cm
\hskip -2cm
\epsfig{file=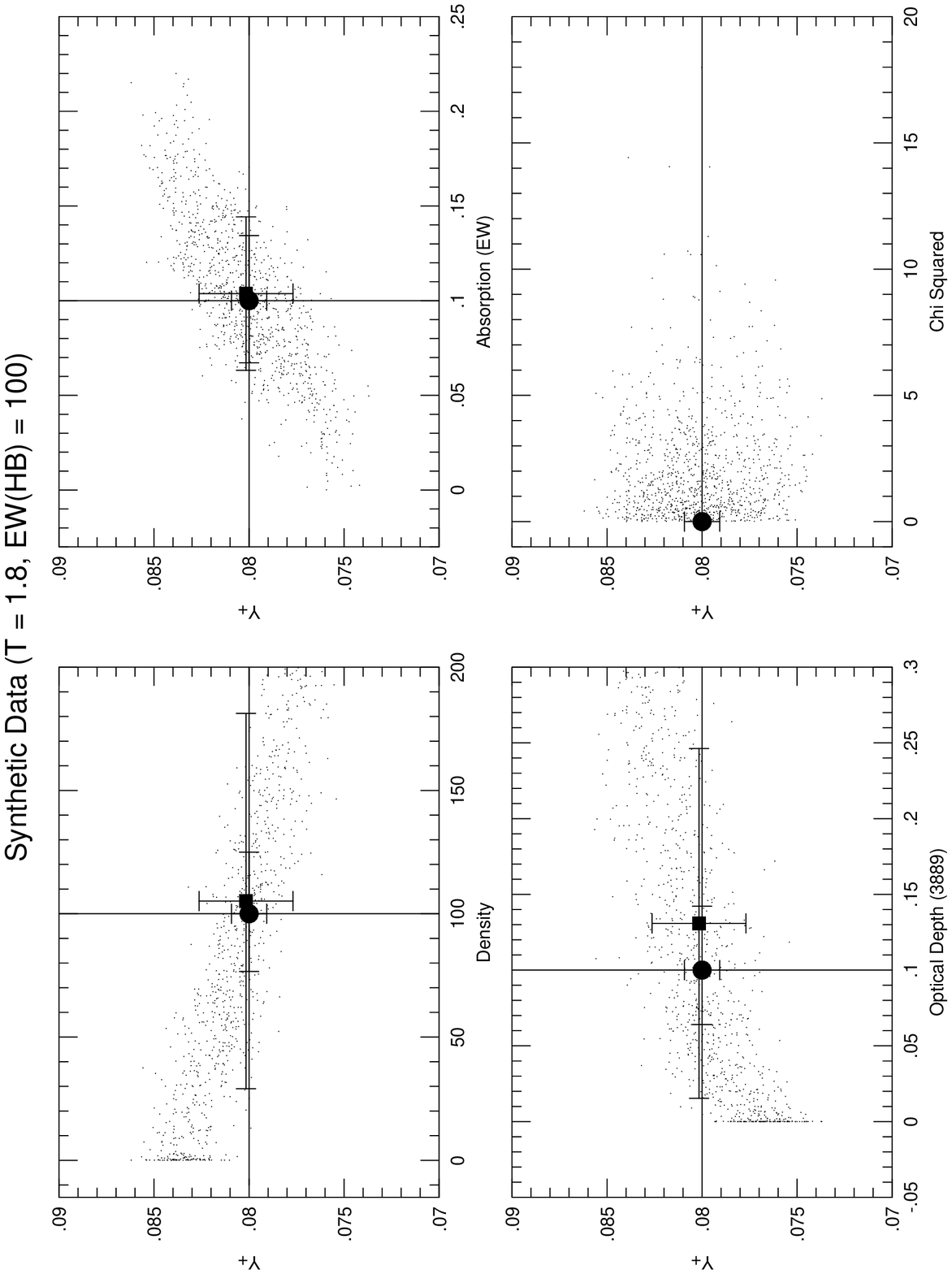}
\end{figure}

\newpage

\begin{figure}
\vskip -2cm
\hskip -2cm
\epsfig{file=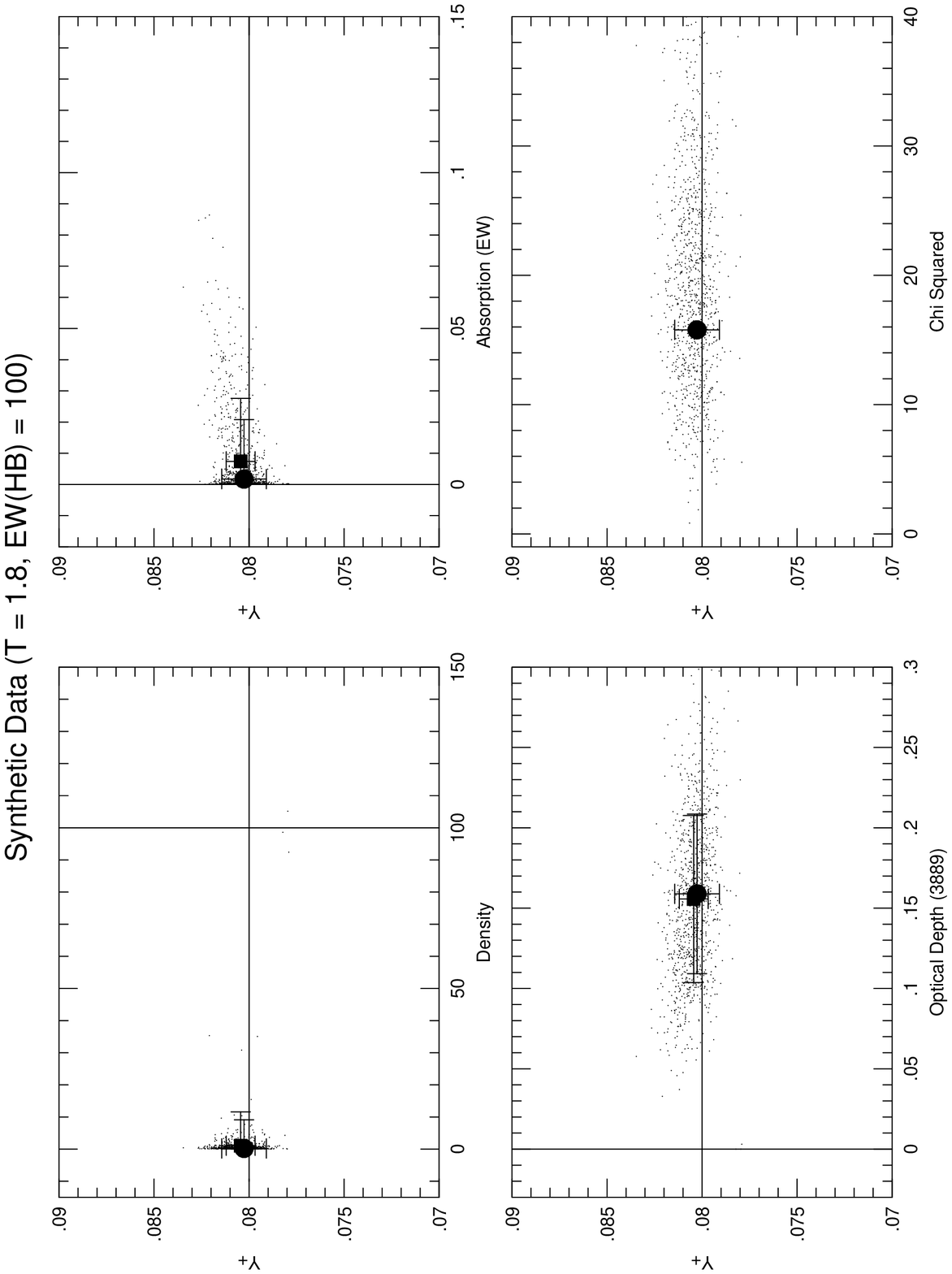}
\end{figure}

\newpage

\begin{figure}
\vskip -2cm
\hskip -2cm
\epsfig{file=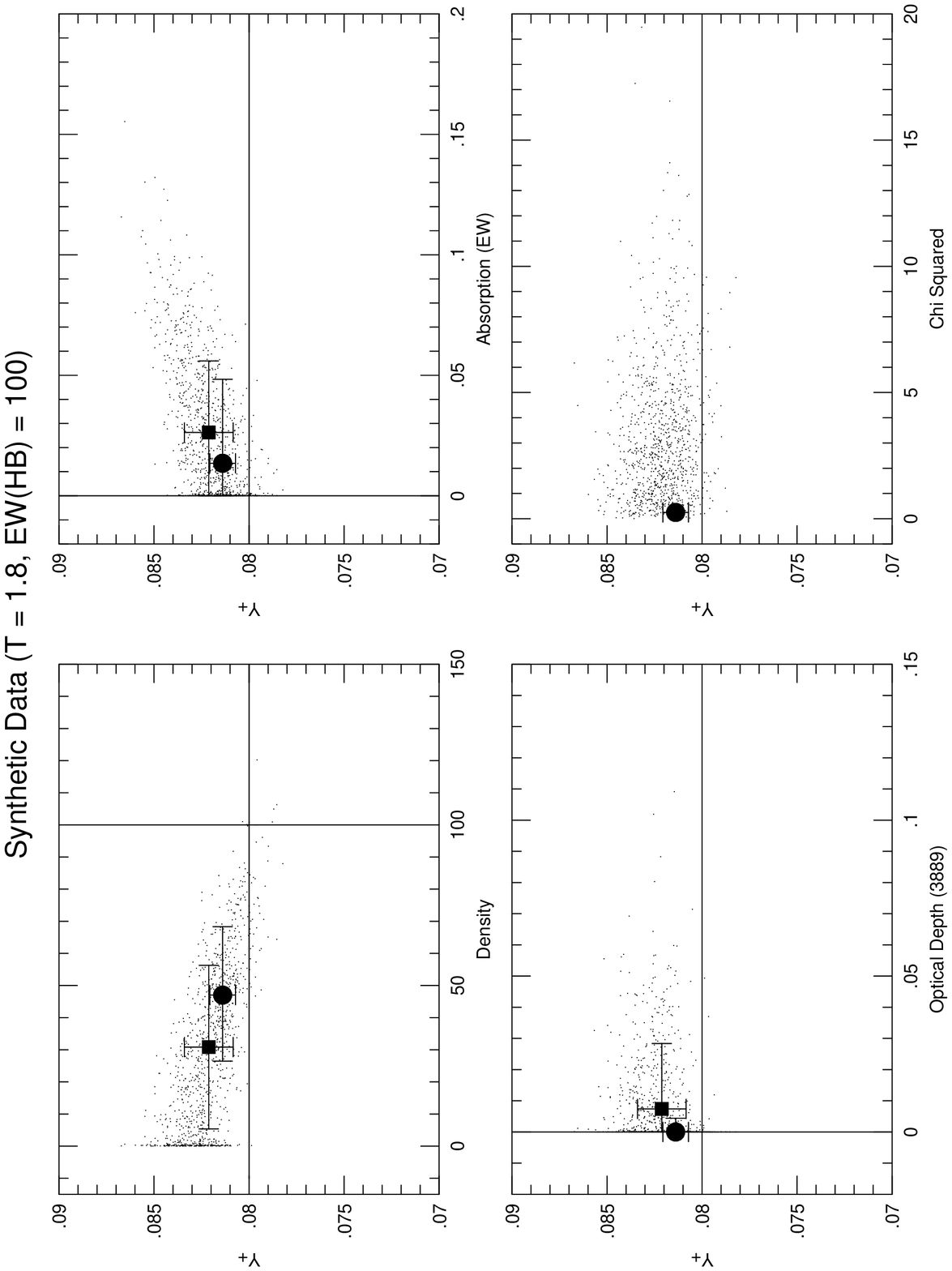}
\end{figure}

\newpage

\begin{figure}
\vskip -2cm
\hskip -2cm
\epsfig{file=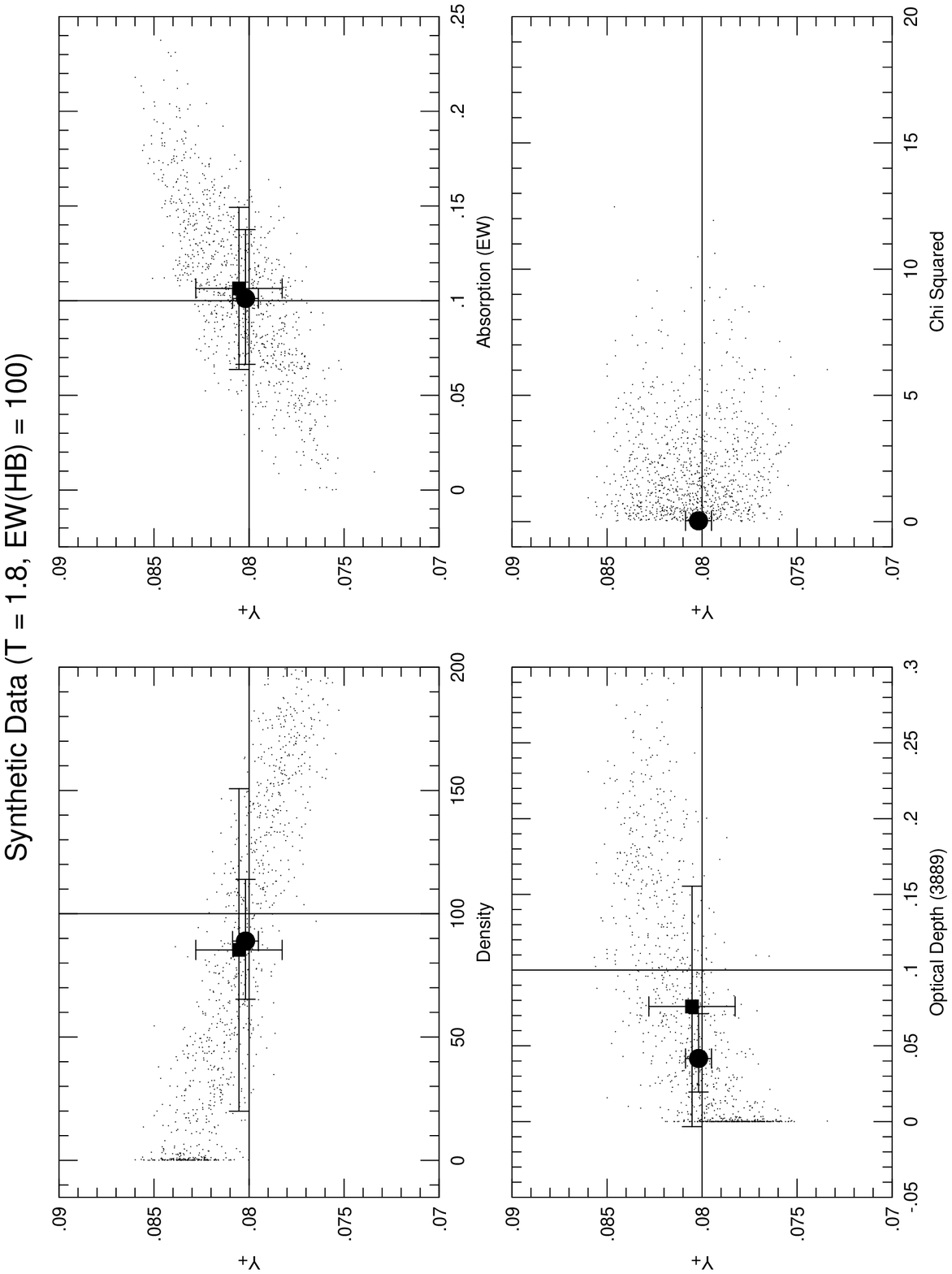}
\end{figure}

\newpage

\begin{figure}
\vskip -2cm
\hskip -2cm
\epsfig{file=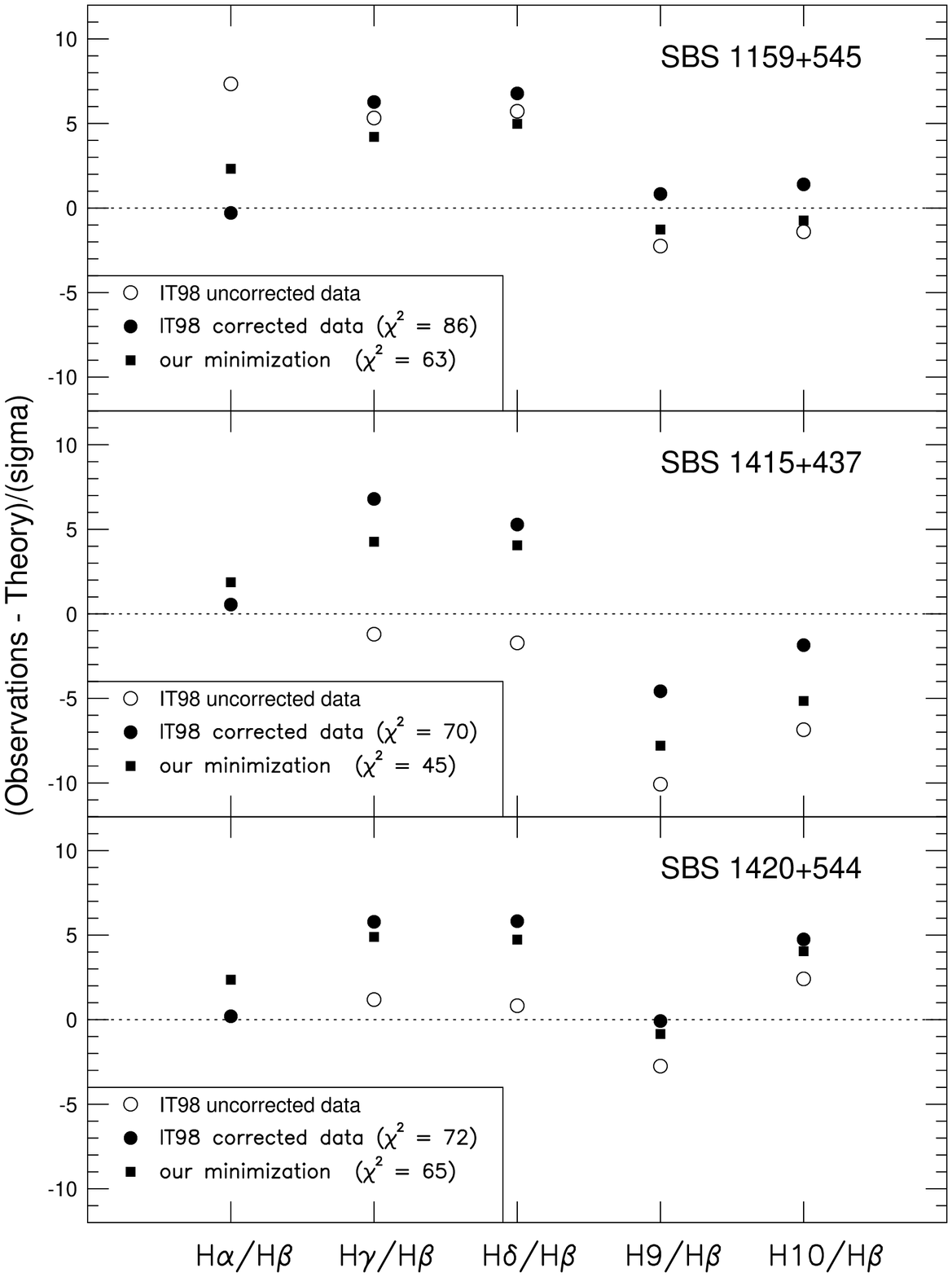}
\end{figure}

\end{document}